\documentclass[useAMS,usenatbib]{mn2e}
\usepackage{graphicx}
\usepackage{subfigure}
\usepackage{longtable}
\usepackage{lscape}
\usepackage{deluxetable}

\newcommand{\ltappeq}{\raisebox{-0.6ex}{$\,\stackrel{\raisebox{-.2ex}{$\textstyle <$}}{\sim}\,$}}

\title[Properties of CVs in H$\alpha$ Surveys]{The Properties of 
Cataclysmic Variables In Photometric H$\alpha$ Surveys}

\author[A.R. Witham et al.]{
A. R. Witham,$^1$
C. Knigge,$^1$
B. T. G\"ansicke,$^2$
A. Aungwerojwit,$^2$
\newauthor
R. L. M. Corradi,$^{3,4}$
J. E. Drew,$^5$
R. Greimel,$^3$
P. J. Groot,$^6$
\newauthor
L. Morales--Rueda,$^6$
E. R. Rodriguez--Flores,$^{4,7}$
P. Rodriguez--Gil,$^4$
\newauthor
and D. Steeghs.$^8$
\\
$^{1}$ School of Physics \& Astronomy, University of Southampton, Highfield, SO17 1BJ, U.K.\\
$^{2}$ Department of Physics, University of Warwick, Coventry CV4 7AL, U.K. \\
$^{3}$ Isaac Newton Group of Telescopes, Apartado de correos 321, E-38700 Santa Cruz de la Palma, Tenerife, Spain \\
$^{4}$ Instituto de Astrofisica de Canarias, 38200 La Laguna, Tenerife, Spain \\
$^{5}$ Imperial College of Science, Technology and Medicine, Blackett Laboratory, Exhibition Road, London, SW7 2AZ, U.K.\\
$^{6}$ Afdeling Sterrenkunde, Radboud Universiteit Nijmegen, Faculteit NWI, Postbus 9010, 6500 GL Nijmegen, the Netherlands \\
$^{7}$ Instituto de Geofisica y Astronomia, Calle 212, No. 2906, CP 11600, La Habana, Cuba \\
$^{8}$ Harvard-Smithsonian Center for Astrophysics, 60 Garden Street, Cambridge, MA 02138, USA \\
}

\begin{document}

\date{2005 August 5}

\pagerange{\pageref{firstpage}--\pageref{lastpage}} \pubyear{2005}

\maketitle

\label{firstpage}

\begin{abstract}

We report on the properties of 71 known cataclysmic variables (CVs) in 
photometric H$\alpha$\ emission line surveys. Our study is motivated by
the fact that the Isaac Newton Telescope (INT) Photometric H$\alpha$
Survey of the northern galactic plane (IPHAS) will soon provide
$r^\prime$, $i^\prime$ and narrow-band H$\alpha$ measurements down to
$r^\prime \simeq 20$ for all northern objects between $-5^o < b <
+5^o$. IPHAS thus provides a unique resource, both for studying the
emission line properties of known CVs and for constructing a new CV
sample selected solely on the basis of H$\alpha$\ excess. Our goal
here is to carry out the first task and prepare the way for the
second. In order to achieve this, we analyze data on 19 CVs
already contained in the IPHAS data base and supplement this with
identical observations of 52 CVs outside the galactic plane. 

Our key results are as follows: (i) the recovery rate of known CVs as
H$\alpha$ emitters in a survey like IPHAS is $\simeq 70$ per cent; (ii) of the $\simeq 30$ per cent of CVs which were not recovered $\simeq 75$ per cent were clearly detected but did not exhibit a significant H$\alpha$ excess at the time of our observations;
(iii) the recovery rate depends only weakly on CV type; (iv) the recovery
rate depends only weakly on orbital period; (v) short-period
dwarf novae tend to have the strongest H$\alpha$\ lines.  These results imply that photometric emission line searches provide an efficient way of constructing CV samples that are not biased against detection of
intrinsically faint, short-period systems. 

\end{abstract}

\begin{keywords}
surveys -- binaries: close -- novae, cataclysmic variables
\end{keywords}

\section{Introduction}

Cataclysmic variables (CVs) are interacting binary systems in which 
a white dwarf (WD) primary accretes matter from a main-sequence
secondary via Roche lobe overflow. If the magnetic field of the WD is
dynamically unimportant, the accretion process takes place entirely
via a disc surrounding the WD. By contrast, if the magnetic field is
very strong, the accretion stream from the secondary is channelled
directly onto the magnetic poles of the WD. Finally, in the
intermediate range of field strengths, a partial disc may form that is
truncated on the inside by the WD magnetic field. 

An observational feature that is common to all of these different
accretion modes are (Balmer) emission lines. In weakly-magnetic CVs, the
observed line emission may originate from optically thin or irradiated
parts of the accretion disc \citep{1980ApJ...235..939W}. In magnetic
CVs, line emission may be produced in the accretion stream, any
residual accretion disc, or in the so-called accretion curtains that
channel material from the inner edge of a truncated disc to the 
magnetic poles of the WD.  The irradiated secondary star in CVs can
also contribute an appreciable contribution to the observed line emission, particularly in polars. \citet{1995cvs..book.....W} provides a
comprehensive review of CVs, including their spectroscopic properties. 

Given the ubiquity of line emission amongst CVs, emission line surveys 
offer a powerful way to find new CVs (for a current example of such a
CV search, see \citealt{2002ASPC..261..190G} and
\citealt{2005A&A...443..995A}). What makes this strategy particularly
promising is that, empirically, the intrinsically 
faintest, low mass transfer rate ($\dot{M}$) systems tend to have the
largest Balmer line 
equivalent widths (EWs; \citealt{1984ApJS...54..443P}). Population
synthesis models suggest that such low accretion rate systems should
totally dominate the galactic CV population and should be found
predominantly at short orbital periods, that is below the well-known
CV ``period gap'' between 2.2 hrs and 2.8 hrs \citep{1993A&A...271..149K,
1997MNRAS.287..929H}. However, this dominant population of faint,
short-period CVs has proven quite elusive, and it is still not clear
whether this is due to selection effects or whether it points to a
serious flaw in our understanding of CV evolution. Since emission line
surveys should be very good at finding low $\dot{M}$\ CVs, they should
be an  excellent way of detecting the  large population of faint,
short-period CVs, if it exists.

The INT/WFC Photometric H$\alpha$ Survey of the northern galactic
plane (IPHAS) is currently surveying the Milky Way in
$r^\prime$,$i^\prime$ and H$\alpha$ and provides an excellent data
base for a detailed CV search at low galactic latitudes.  The survey
goes to a depth of $r^\prime~\simeq~20$\ and covers the latitude range
$-5^o < b < +5^o$. A detailed introduction to the survey is given by 
\citet{2005MNRAS.362..753D}.

In this paper, we investigate the properties of the known population of CVs
that have been observed by IPHAS, supplemented with additional
observations of CVs outside the galactic plane with the IPHAS
set-up. Such a study is useful for two main reasons. First, it allows
us to determine the recovery rate of CVs as  
H$\alpha$ emitters as a function of CV sub-class, orbital period and
apparent magnitude. These recovery rates effectively specify the
completeness of the survey with respect to CVs, which is crucial for
the interpretation of the new CV sample that is currently being
constructed based on IPHAS data. Second, the existence of a large,
uniform sample of broad- and narrow-band magnitudes for known CVs
allows us to check for and update correlations between key observables 
(for example H$\alpha$\ excess) and intrinsic CV parameters (for example orbital
period and absolute magnitude).  

The structure of this paper is as follows. Section~\,\ref{obs}
discusses the observations obtained. Section~\,\ref{samples} 
discusses the two CV samples analysed in the paper and how they have
been observed. The techniques for selecting CVs from photometric
data are presented in section~\,\ref{cv_sel}.  In
section~\,\ref{res_s}, we present and discuss our estimates for CV
recovery/detection rates with IPHAS-like emission line
surveys. In section~\,\ref{correl}, we consider correlations between key
observables and system parameters. Our findings are discussed in
section~\,\ref{discuss}, focusing particularly on the ability of IPHAS
to uncover the long-sought population of faint, short-period CVs. We
summarize our conclusions in section~\,\ref{conc}.

\section{Observations}
\label{obs}

Our study relies on two observational data sets. The first is composed
of the IPHAS galactic plane observations that were completed up to and
including July 2004; this amounts to about 1/3 of the final survey
area. The second consists of a series of imaging observations obtained
in June 2004 of northern CVs with galactic latitudes $|b| > 5^o$. Both
datasets were obtained with the same instrumental set-up and
observing strategy, and both were reduced with the standard IPHAS pipeline.  

Full details of the IPHAS observing strategy and reduction procedures
may be found in \citet{2005MNRAS.362..753D}. Briefly, all observations
were obtained using the Wide Field Camera (WFC) on the 
Isaac Newton Telescope, which gives a spatial pixel size of $0.333$\arcsec x
$0.333$\arcsec\ over a field of view of approximately 0.3 square
degrees. Photometry was carried out using a set of three
filters, comprising a narrowband H$\alpha$\ filter and an additional
set of two broadband Sloan $r^\prime$\ and $i^\prime$\
filters. Exposure times were 120s for the H$\alpha$\ images, 10s for
the $i^\prime$ band images, and 10s [30s] 
seconds for all $r^\prime$ band images obtained before [after] June 2004. 
Finding charts from \citet{2001PASP..113..764D} were used to identify
the CVs in our images. 

The final IPHAS data base will contain only observations that meet a
strict set of quality constraints. However, in order to avoid
restricting our CV sample too much in the present analysis, we did 
not limit ourselves to fields that have passed the IPHAS quality
controls. Instead, for each CV, we simply took the data from the best
available set of observations.\footnote{Note that a given CV can be
present in multiple IPHAS pointings because (i) most locations are
observed at least twice with the IPHAS tiling pattern, and (ii) poor
quality observations may already have been repeated.}. Thus some of
the data analysed here will not be included in the final published survey 
catalogues. Instances in which the data quality of a field was below the nominal
limits will be indicated explicitly.  However, our results from these fields are
consistent with the results from fields with data quality that passes the IPHAS quality controls.

We finally note that, in all of our analysis below, we only work with
sources brighter than a pre-defined magnitude limit of $r^\prime =
19.5$. In principle, IPHAS goes somewhat deeper than this, to about
$r^\prime \simeq 20$. However, the increased photometric scatter
exhibited by the very faintest sources makes it counter-productive to
include these sources in our selection algorithm (see
Section~\ref{cv_sel}).

\section{Construction of CV Samples}
\label{samples}

As noted above, the data we use in our study come from two distinct
sources. The first is the standard IPHAS database, the second is a 
set of observations of CVs at higher galactic latitudes observed with
the IPHAS set-up. The majority of the fields in the second 
data set exhibit a significantly lower object density than the IPHAS
fields. This may be expected to affect CV recovery rates, and we
therefore retain the distinction between In-Plane and Off-Plane
samples in our study. 

\subsection{The In-Plane CV Sample}
\label{in-pl_sample}
Adopting a limiting magnitude of $r^\prime = 19.5$ the IPHAS imaging database as of July 2004 contains 19 known cataclysmic variables\footnote{Note that we omit the double-degenerate
AM CVn systems from consideration; these He-rich systems belong to a
different evolutionary channel and are not expected to show H$\alpha$
emission.} listed in the Ritter \& Kolb
catalogue \citep{2003A&A...404..301R}\footnote{The Ritter \& Kolb catalogue is available online
at http://physics.open.ac.uk/RKcat.}. 
We extracted standard IPHAS catalogues, containing object
positions and magnitudes, for all of the IPHAS fields that include
these CVs.  As noted above, when multiple observations were available
for a given CV, we always worked with data from the field observed
under the best conditions. Information and data on each CV in the sample are listed
in Table~\ref{tab1_b}. This includes a flag to mark data that were
extracted from fields that do not meet the nominal IPHAS quality
thresholds. 

\subsection{The Off-Plane CV Sample}
The targets for off-galactic-plane ($|b| > 5^o$) observations were
again chosen from \citet{2003A&A...404..301R}, with the main
additional constraint being that the target should be observable in 
June 2004. A sample of 52 Off-Plane CVs turned out to satisfy our magnitude limit of $r^\prime <19.5$. The IPHAS pipeline-reduced data were extracted in exactly
the same way as for the In-Plane CVs.

One target, the dwarf nova (DN) IY UMa, deserves additional 
comment. It was found to have a faint close companion which
is not identified in the finding charts and has not 
been previously described in literature. In the $r^\prime$ band, the
faint companion 
has a magnitude of 19.27 mag and is separated from IY UMa by 3.5
arcsec. The faint companion has resulted in IY UMa being detected as
an unresolved blend in the $i^\prime$ band, which makes the $i^\prime$ 
band magnitude unreliable in this case.

\section{Selecting H$\alpha$ Excess Objects}
\label{cv_sel}

\begin{figure*}
\includegraphics[angle=-90,width=\textwidth]{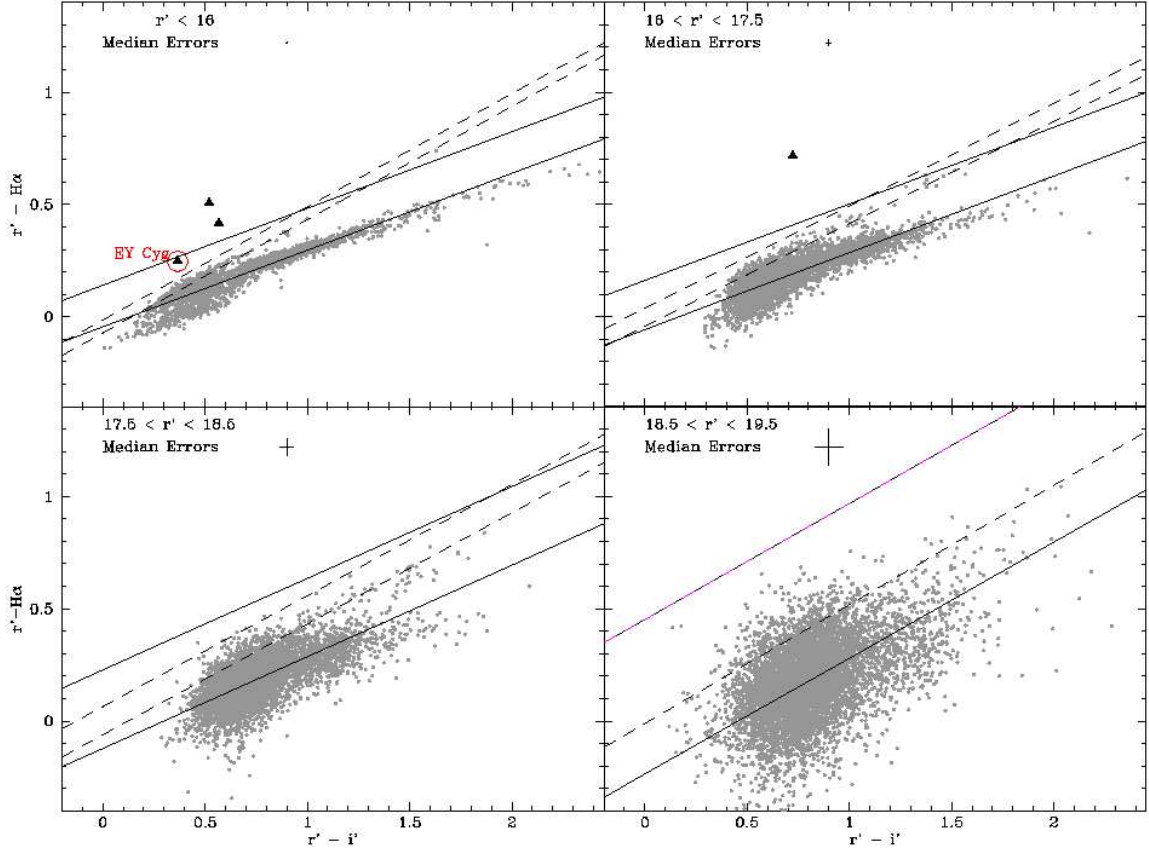}
\caption{\label{selection} An illustration of the selection criteria
used to identify strong emission line objects on colour-colour
diagrams. The data shown here are all from the IPHAS field 5455 which surrounds the bright
dwarf nova EY Cyg. First, the data are split up into four magnitude
bins, as shown in the four panels. Objects with H$\alpha$ excess
should be located near the top of a colour-colour diagram. The thick lines
illustrate the fits to the data, and the thin lines illustrate the
corresponding cuts used to single out clear H$\alpha$ emitters. The
solid lines show the original least squares fit to the full dataset,
along with the corresponding selection cut.  The dashed lines show the
final fit and cut lines, obtained by applying an iterative
$\sigma$-clipping techniques to the initial fit. The goal of the
$\sigma$-clipping is to fit only the the upper
boundary of the main stellar locus in
the colour-colour diagram. Objects selected as H$\alpha$ emitters are
shown as large triangles. Note that the cut lines shown here are only
approximate, as the actual selection criterion used also considers the
errors on each individual datapoint (see Equation~\ref{sel}).}
\end{figure*}

We have developed a systematic selection algorithm to identify 
H$\alpha$ excess outliers (and hence likely emission line objects)
from $r^\prime~-~$H$\alpha$~vs~$r^\prime~-~i^\prime$ colour-colour
plots of individual fields. The selection process does not require
user intervention and 
allows us to separate the CVs into those that show clear H$\alpha$
emission and those that do not. We can then use this information to
test for statistical completeness and selection biases in samples of
CVs constructed solely on the basis of photometric H$\alpha$\
excess. In the following subsections, we describe the sequence of
steps used by our selection algorithm to identify likely emission line
sources.

\subsection{Initial Selection Cuts}
\label{initial_selection}

The first step in our selection procedure is to apply the following
criteria to all objects in the catalog for each field in the IPHAS
database. First, objects must be detected in all three photometric
bands and must not fall on bad pixels of the CCDs.  Second, the objects must
have positions in the three photometric bands that match to
within 1 arcsec.  Third, objects must be classified as ``stellar'' in the
$i^\prime$ band, and ``stellar'' or ``probably stellar'' in the $r^\prime$ and
H$\alpha$ bands. We allow ``probably stellar'' classification in $r^\prime$
and H$\alpha$ to avoid excluding stars surrounded by extended
H$\alpha$ nebular emission. This prevents the removal of CVs with nova
shells (for example T Aur in our In-Plane sample).  

\subsection{Locating the Main Stellar Locus}

Objects surviving the initial cuts are then subjected to an iterative
selection algorithm (illustrated in  Fig.\,\ref{selection}). First,
they are divided into four magnitude bins, based on their $r^\prime$
band magnitudes. Next
$r^\prime~-~$H$\alpha$~versus~$r^\prime~-~i^\prime$ colour-colour
plots are created for each magnitude bin in each 
field. 

We next carry out an initial straight line least squares fit to the objects in
each magnitude bin. Objects with strong H$\alpha$\ emission lines
should exhibit an excess in $r^\prime~-~$H$\alpha$\, and will
therefore generally be located above the main stellar loci 
in the plots.  The least squares fit is an initial attempt to fit the
main stellar loci, and so we expect 
H$\alpha$ emitters to reside above the fit in the plot. For fields
with low number density, which includes 
the majority of fields outside the galactic plane, this initial fit to
all objects often works well. However, in fields with higher number
densities, the stellar locus often splits due to the presence of
different stellar populations (giants and dwarfs) and different
reddening values (see \citealt{2005MNRAS.362..753D} for a
detailed explanation). The key point for our purposes is that the
upper locus of points in the 
colour-colour plots often runs above the initial least
squares fit in fields dominated by more distant/highly
reddened population. Thus a selection criterion based solely on an
object's position relative to our initial fit would result
in the erroneous selection of many ordinary late-type unreddened
main-sequence stars which reside in the upper locus of points.

\subsection{Identifying The Upper Boundary of the Main Stellar Locus}

In order to prevent the selection of unreddened objects with no
H$\alpha$ emission, we use an iterative $\sigma$-clipping technique
to force the fit onto the upper 
boundary of the main stellar locus. The fact that the unreddened
objects reside at the top of the main stellar loci is used in our
technique to exclude gradually points below the initial fit line in
subsequent iterations. In practice, we use four iterations to arrive
at a final best-bet fit to the the upper 
boundary of the main stellar locus. The final iteration is used to
give precedence to objects that are well separated from the initial
fit (within reasonable limits) rather than those objects that lie in
the regions where the upper 
boundary of the main stellar locus is no longer clearly separated from
the lower regions of the main stellar locus.  

However, this final fit may not always be appropriate, such as in
fields where the stellar locus is not actually split. In these cases,
the gradient of the upper
boundary of the main stellar locus fit often matches closely
with the initial least squares fit or exhibits a nonphysically
large slope. If either of these conditions is met, the 
selection of H$\alpha$ emitters is based on the initial fit; in all other cases, the final
fit is used to select emitters. Empirically, we find that this method
gives reasonable results 94 per cent of the time (compared to visual
inspection). That is, the fit runs through the upper boundary of the
stellar locus when it is visible and otherwise fits the whole stellar
locus in a satisfactory manner. 

\subsection{Selecting H$\alpha$ Excess Objects}  

Once the appropriate fit for each bin has been decided, we identify
objects significantly above the chosen fit as likely H$\alpha$\
emitters. Our selection criterion takes into account both the scatter
of points around the stellar loci and the errors on the colours of
each individual datapoint. We first define the H$\alpha$\ excess for
each object as 
\begin{equation}
\Delta \rmn{H}\alpha = (r^\prime-\rmn{H}\alpha)_{\rmn{obs}} - (r^\prime-\rmn{H}\alpha)_{\rmn{fit}}
\end{equation}
where $(r^\prime-$H$\alpha)_{\rmn{obs}}$ is the observed value of the colour and
$(r^\prime-$H$\alpha)_{\rmn{fit}}$ is the value obtained from the fit for the
corresponding observed value of $r^\prime-i^\prime$. The selection criterion is then
given in terms of H$\alpha$\ excess by 
\begin{equation}
\Delta \rmn{H}\alpha > C\sqrt{[rms^2 + \sigma_{(r^\prime-\rmn{H}\alpha)}^2 + m_{\rmn{fit}}^2\sigma_{(r^\prime-i^\prime)}^2]}.
\label{sel}
\end{equation}
Here $rms$ is the root mean square value of the residuals
around the fit, $C$ is a constant (we use values of 4.5 for our
initial fits and 5 for the fits with $\sigma$-clipping), 
$m_{\rmn{fit}}$ is the gradient of the fit line, and $\sigma_{(r^\prime-H\alpha)}$
and $\sigma_{(r^\prime-i^\prime)}$ are the errors on the observed colours.

Note that the cut lines shown here are only
approximate, as the actual selection criterion used also considers the
errors on each individual datapoint (see Equation~\ref{sel}).

In the example shown in Fig.\,\ref{selection}, the lines illustrate the
initial and final selection cuts applied. However the cut lines shown are only
approximate, as the actual selection criterion used also considers the
errors on each individual datapoint (see Equation~\ref{sel}) and this is not reflected
by the cut lines shown in the figure.  This explains why in some
cases objects appearing above the illustrated cuts  
are not selected as emitters.  The selected H$\alpha$ emitters
are indicated by the large triangles.  The figure shows data from
IPHAS field 5455, which contains the CV EY Cyg. Note that the data
quality of this field falls below the IPHAS standards, that is this field
will be re-observed for the final IPHAS survey at a latter date. EY 
Cyg is a bright dwarf nova 
with V-band magnitudes of 11.4 and 15.5 in outburst and quiescence, 
respectively. The IPHAS photometry ($r^\prime$ = 14.1) suggests the system was
observed near quiescence, when it is known to display clear H$\alpha$
emission \citep{1997A&AS..122..495M}. The CV is marked by a large open
circle and was correctly selected as a likely emission line object by
our algorithm. We note that, in this case, the final fit succeeded in
identifying EY Cyg as an emission line object, whereas a selection
based on the initial fit to all the data would have failed to do so.

It is worth emphasizing that we make no claim that the method
presented here is in any strict sense optimal for selecting H$\alpha$
emitters. The design goals for our algorithm are merely that it should
be simple, fully automated and produce reasonable results in the vast
majority of cases (when compared to a fully interactive outlier
selection). Consequently, the selection criteria we use are 
essentially empirical and tailored to provide adequate results in most
situations.

\begin{table*}
\begin{minipage}{160mm}
\caption{\label{tab1_a} Summary of the existing data for the CVs in
the two samples.   Magnitude ranges are drawn
directly from the Downes \textit{et al} catalogue.  Orbital period
information is drawn directly from the Ritter and Kolb
catalogue.  CV classification data are adapted from the information
contained  within CVCat.}
\setlength{\tabcolsep}{1.1ex}
\begin{tabular}{llllcl}
\hline
GCVS Name            	& Other name	      		& Magnitude Range	& Orbital   			& CV				& CV				\\	
	             	&				&			& Period\tablenotemark{a}	& Flag\tablenotemark{b}		& Type \tablenotemark{c}	\\	
	             	&				&			& (days)			&				&				\\	
\hline
\multicolumn{6}{l}{\textbf{Off-Plane}} \\
\hline
RZ LMi                      	& PG 0948+344         		& 14.2 V   -   16.8 V	& 0.058500*	& 0 		& DN/SU  				\\	
RU LMi                      	& CBS-119/Ton 1143    		& 13.8 p -  19.5 p   	& 0.251000 	& 0 		& DN     				\\	
CH UMa                      	& PG 1003+678         		& 10.7 V -  15.3 V   	& 0.343184 	& 0 		& DN     				\\	
SW Sex                      	& PG 1012-029         		& 14.8 B -  16.7 B   	& 0.134938 	& 0 		& NL/SW  				\\	
GG Leo                      	& RX J1015.5+0904     		& 16.5 V -  18.8 V   	& 0.055471 	& 0 		& AM     				\\	
CP Dra                      	&                     		& 14.3 p -  20   p   	& 0.081600*	& 0 		& DN     				\\	
CI UMa                      	& SVS 1755            		& 13.8 p -  18.8 V   	& 0.060000 	& 0 		& DN/SU  				\\	
KS UMa                      	& RX J1020+5304       		& 13.0 p -  17.0 p   	& 0.068000 	& 0 		& DN     				\\	
FIRST J102347+003841        	&                     		& 17.3 V -  17.8 V   	& 0.198115 	& -1		& -      				\\	
U Leo                       	& BD +14 2239         		& 10.5 v - $<$15   v   	& 0.267400:	& 0 		& N:     				\\	
DW UMa                      	& PG 1030+590         		& 14.9 V -  18.  V   	& 0.136607 	& 0 		& NL/VY  				\\	
DO Leo                      	& PG 1038+155         		& 16.0 B -  17.0 B   	& 0.234515 	& 0 		& NL     				\\	
IY UMa                      	& Tmz V85             		& 13.0 p -  18.4 p   	& 0.073909 	& 0 		& DN/SU  				\\	
CW 1045+525                 	&                     		& 16.5 B -           	& 0.271278 	& 0 		& DN     				\\	
SX LMi                      	& CBS-31/Ton 45       		& 13   V -  17.4 V   	& 0.067200 	& 0 		& DN/SU  				\\	
CY UMa                      	& SVS 2198            		& 12.3 V -  17.8 V   	& 0.067950 	& 0 		& DN/SU  				\\	
AN UMa                      	& PG 1101+453         		& 13.8 B -  20.2 B   	& 0.079753 	& 0 		& AM     				\\	
ST LMi                      	& CW 1103+254         		& 15.0 V -  17.2 V   	& 0.079089 	& 0 		& AM     				\\	
AR UMa                      	& 1ES 1113+432        		& 13.3 V -  16.5 V   	& 0.080501 	& 0 		& AM     				\\	
DP Leo                      	& 1E 1114+182         		& 17.5 B -  19.5 B   	& 0.062363 	& 0 		& AM     				\\	
TT Crt                      	& FSV 1132-11         		& 12.7 V -  16.3 V   	& 0.268420 	& 0 		& DN     				\\	
RZ Leo                      	&                     		& 11.5 V -  19.2 V   	& 0.076038 	& 0 		& DN/SU  				\\	
T Leo                       	& BD +4 2506a         		& 10.0 V -  15.9 V   	& 0.058820 	& 0 		& DN/SU  				\\	
DO Dra                      	& YY Dra?             		& 10.6 B -  16.7 B   	& 0.165374 	& 0 		& IP     				\\	
TW Vir                      	& PG 1142-041         		& 12.1 v -  16.3 v   	& 0.182670 	& 0 		& DN     				\\	
EU UMa                      	& RE 1149+28          		& 16.5 B -  16.8 B   	& 0.062600 	& 0 		& AM     				\\	
BC UMa                      	& GR 102              		& 10.9 B -  18.3 B   	& 0.062610 	& 0 		& DN/SU  				\\	
IR Com                      	& S 10932             		& 13.5 B -  18.  B   	& 0.087039 	& 0 		& DN:/SU:				\\	
EV UMa                      	& RX J1307+53         		& 17.1 V -  20.8 V   	& 0.055338 	& 0 		& AM     				\\	
HV Vir                      	& NSV 6201            		& 11.5 V -  19.0 V   	& 0.057069 	& 0 		& DN/SU  				\\	
HS Vir                      	& PG 1341-079         		& 13.0 B -  16.6 V   	& 0.076900 	& 0 		& DN     				\\	
OU Vir                      	& 1432-0033           		& 14.5 p -  18.5 j   	& 0.072730 	& 0 		& DN/SU  				\\	
UZ Boo                      	& HV 10426            		& 11.5 v -  20.4 V   	& 0.125000:	& 0 		& DN/SU  				\\	
KV Dra                      	& RX J1450.5+6403     		& 11.8 V -  17.1 V   	& 0.058760 	& 0 		& DN/SU  				\\	
TT Boo                      	& HV 3681             		& 12.7 v -  19.2 V   	& 0.077000:	& 0 		& DN/SU  				\\	
NY Ser                      	& NSV 6990            		& 14.8 V -  17.9 V   	& 0.097800 	& 0 		& DN/SU  				\\	
M5 V101                     	&                     		& 17.5 p -  20.9 V   	& 0.242000 	& 0 		& DN     				\\	
ES Dra                      	& PG 1524+622         		& 13.9 p -  16.3 p   	& 0.179000 	& 0 		& DN/SU  				\\	
QW Ser                      	& Tmz V46             		& 12.8 p - $<$15.3 p   	& 0.074570 	& 0 		& DN     				\\	
ASAS 153616-0839.1          	&                     		& 11.5 V - $<$15.4 V   	& 0.063600*	& 0 		& DN     				\\	
LX Ser                      	& Stepanian's star    		& 13.3 B -  17.4 B   	& 0.158432 	& 0 		& NL/SW  				\\	
CT Ser                      	& N Ser 1948 (r)      		& 7.9 v -  16.6 p    	& 0.195000 	& 0 		& N      				\\	
SS UMi                      	& PG 1551+719         		& 12.6 V -  17.6 V   	& 0.067780 	& 0 		& DN/SU  				\\	
MR Ser                      	& PG 1550+191         		& 14.9 V -  17.  V   	& 0.078798 	& 0 		& AM     				\\	
\footnotetext[1]{A colon following the orbital period signifies the value is uncertain, an asterisk signifies the value has been estimated from the known superhump period.}
\footnotetext[2]{Possible entries in this column; 0 = good, 1= uncertain, -1 = not in CV Cat, -2 = non cv.}
\footnotetext[3]{Possible entries in this column; DN = dwarf nova, N = nova, NL = novalike variable, NR = recurrent nova, AM = polar (AM Her), IP = intermediate polar, - = non classified.  Any entries in this column following a forward slash are the secondary classification of the CV, possible additional entries are; SU = SU UMa system, VY = VY Scl system, ZC = Z Cam system, UG = U Gem system, SW = SW Sex system, Na = fast nova, Nb = slow nova.  A colon following any of these entries signifies the classification is uncertain.}
\end{tabular}
\end{minipage}
\end{table*}

\begin{table*}
\begin{minipage}{160mm}
\contcaption{}
\setlength{\tabcolsep}{1.1ex}
\begin{tabular}{llllcl}
\hline
GCVS Name            	& Other name	      		& Magnitude Range	& Orbital   			& CV				& CV				\\	
	             	&				&			& Period\tablenotemark{a}	& Flag\tablenotemark{b}		& Type \tablenotemark{c}	\\	
	             	&				&			& (days)			&				&				\\	
\hline
\multicolumn{6}{l}{\textbf{Off-Plane}} \\
\hline
RX J1554.2+2721             	&                     		& 16.8 B -           	& 0.105462 	& 0 		& AM     				\\	
QZ Ser                      	& HadV4               		& 12.7 p - $<$14.9 p   	& 0.083161 	& 0 		& DN     				\\	
VW CrB                      	& Antipin V21         		& 14.0 p - $<$17.5 b   	& 0.070700*	& 0 		& DN     				\\	
1RXS J161008+035222         	&                     		&                    	& 0.132200 	& 0 		& AM     				\\	
X Ser                       	& HV 3137             		& 8.9 p -  18.3 p    	& 1.480000 	& 0 		& N:/Nb: 				\\	
V589 Her                    	& S 10296             		& 14.1 p - $<$17.5 p   	& 0.090500*	& 0 		& DN     				\\	
V844 Her                    	& Antipin V43         		& 12.5 p -  17.5 p   	& 0.054643 	& 0 		& DN/SU  				\\	
V699 Oph                    	& HV 10577            		& 13.8 p -  18.5 p   	& 0.068300*	& -2		& -      				\\	
RW UMi                      	& SVS 1359            		& 6   p -  18.5 B    	& 0.059100 	& 0 		& N/Nb   				\\	
V841 Oph                    	& BD -12 4633         		& 4.2 v -  13.5 v    	& 0.601400 	& 0 		& N/Nb   				\\	
\hline                                                                                                                                                 
\multicolumn{6}{l}{} \\                                                                                                                               
\multicolumn{6}{l}{\textbf{Plane}} \\
\hline
   HT Cas                   	&            S 3343   		& 10.8 v - 18.4 v    	& 0.073647 	& 0 		& DN/SU  				\\	
    V Per                   	&       BD +56 406a   		& 9.2 p  - 18.5 p    	& 0.107126 	& 0 		& N      				\\	
   UV Per                   	&        AN 87.1911   		& 11.7 V - 17.9 V    	& 0.064900 	& 0 		& DN/SU  				\\	
   TZ Per                   	&        AN 28.1912   		& 12.3 v - 15.6 v    	& 0.262906 	& 0 		& DN/ZC  				\\	
    T Aur                   	&       BD +30 923a   		& 4.2 p  - 15.2 p    	& 0.204378 	& 0 		& N/Nb   				\\	
   FS Aur                   	&            S 3946   		& 14.4 v - 16.2 v    	& 0.059500 	& 0 		& DN     				\\	
   CW Mon                   	&        AN 61.1936   		& 11.9 v - 16.3 v    	& 0.176600 	& 0 		& DN     				\\	
 V603 Aql                   	&         HD 174107   		& -1.1 v - 11.8 B    	& 0.138500 	& 0 		& N/Na   				\\	   
   CI Aql                   	&        AN 23.1925   		& 8.8 V  - 15.6 p    	& 0.618363 	& 0 		& NR     				\\	
 V446 Her                   	&                     		& 3.0 p  - 17.8 B    	& 0.207000 	& 0 		& N/Na   				\\	
V1425 Aql                   	&                     		& 7.5 v  - $<$19.0 V   	& 0.255800 	& 0 		& N/IP:  				\\	
   EM Cyg                   	&       AN 185.1928   		& 12.5 v - 14.4 v    	& 0.290909 	& 0 		& DN/ZC  				\\	
RX J1951.7+3716             	&                     		& 15.5 B -           	& 0.493000 	& 0 		& -      				\\	   
   EY Cyg                   	&       AN 200.1928   		& 11.4 v - 15.5 v    	& 0.459326 	& 0 		& DN/UG  				\\	
V2306 Cyg                   	& 1WGA J1958.2+3232   		& 15.7 V -           	& 0.181200 	& 0 		& IP     				\\	
 V550 Cyg                   	&            S 3847   		& 14.2 p - $<$17.0 p   	& 0.067200*	& 0 		& DN/UG  				\\	
 V503 Cyg                   	&            S 4524   		& 13.4 V - 17.6 V    	& 0.077700 	& 0 		& DN/SU  				\\	
 V751 Cyg                   	&          SVS 1202   		& 13.2 V - 16.0 V    	& 0.144464 	& 0 		& NL/VY  				\\	
V2275 Cyg                   	& N Cyg 2001, No. 2   		& 6.6 p  -           	& 0.314500 	& 0 		& N      				\\	
\footnotetext[1]{A colon following the orbital period signifies the value is uncertain, an asterisk signifies the value has been estimated from the known superhump period.}
\footnotetext[2]{Possible entries in this column; 0 = good, 1= uncertain, -1 = not in CV Cat, -2 = non cv.}
\footnotetext[3]{Possible entries in this column; DN = dwarf nova, N = nova, NL = novalike variable, NR = recurrent nova, AM = polar (AM Her), IP = intermediate polar, - = non classified.  Any entries in this column following a forward slash are the secondary classification of the CV, possible additional entries are; SU = SU UMa system, VY = VY Scl system, ZC = Z Cam system, UG = U Gem system, SW = SW Sex system, Na = fast nova, Nb = slow nova.  A colon following any of these entries signifies the classification is uncertain.}
\end{tabular}
\end{minipage}
\end{table*}

\scriptsize
\onecolumn
\begin{landscape}
\setlength{\tabcolsep}{1.1ex}
\begin{longtable}{llllccccccccccccccl}
\caption{\label{tab1_b} Summary of the IPHAS data for the CVs in the two samples.} \\
\hline \\[-2ex]
GCVS Name    	& Other name	      	&	RA	&	DEC	& $\ell$	& \textit{b}	& $r^\prime$	& $r^\prime$ 	& $i^\prime$	& $i^\prime$	& \textit{H$\alpha$}	& \textit{H$\alpha$}	& Emitter\tablenotemark{a}		& Initial\tablenotemark{b}		& Field\tablenotemark{c}		& $i^\prime$ band	& $i^\prime$ band	& $i^\prime$ band	& EW\tablenotemark{d}	\\
		&			&		& 		& 		&		&		& error		& 		& error		&			& error			&					& Selection				& Quality 				& Seeing		& Ellipticity		& Magnitude		& (\AA)			\\
		&			&	(2000)	&	(2000)	& ($^{\circ}$)	&  ($^{\circ}$)	&		& 		&		&		&			&			&					& Flag					& Flag					& (arcsec)		&			& Limit			&			\\
\hline
\endfirsthead

\multicolumn{17}{c}{{\tablename} \thetable{} -- Continued} \\[0.5ex]
\hline \\[-2ex]
GCVS Name    	& Other name	      	&	RA	&	DEC	& $\ell$	& \textit{b}	& $r^\prime$	& $r^\prime$ 	& $i^\prime$	& $i^\prime$	& \textit{H$\alpha$}	& \textit{H$\alpha$}	& Emitter\tablenotemark{a}		& Initial\tablenotemark{b}		& Field\tablenotemark{c}		& $i^\prime$ band	& $i^\prime$ band	& $i^\prime$ band	& EW\tablenotemark{d}	\\
		&			&		& 		& 		&		&		& error		& 		& error		&			& error			&					& Selection				& Quality 				& Seeing		& Ellipticity		& Magnitude		& (\AA)			\\
		&			&	(2000)	&	(2000)	& ($^{\circ}$)	&  ($^{\circ}$)	&		& 		&		&		&			&			&					& Flag					& Flag					& (arcsec)		&			& Limit			&			\\
\hline
\endhead

\multicolumn{17}{l}{{Continued on Next Page\ldots}} \\
\endfoot

\\[-1.8ex] \hline
\endlastfoot

\multicolumn{10}{l}{} \\ 
\multicolumn{10}{l}{\textbf{Off-Plane}} \\
\hline
 RZ LMi                      	& PG 0948+344         		& 09:51:48.93	& +34:07:23.9	& 191.29	&  51.04	& 14.789 	& 0.001		& 14.783	& 0.003		& 14.857		& 0.003			& 0			& 0			& 0	& 1.83	& 0.09	& 20.03	&      -5	\\
 RU LMi                      	& CBS-119/Ton 1143    		& 10:02:07.47	& +33:51:00.5	& 191.83	&  53.17	& 17.626 	& 0.006		& 17.296	& 0.013		& 17.007		& 0.008			& 1			& 0			& 0	& 1.42	& 0.10	& 20.47	&      65	\\
 CH UMa                      	& PG 1003+678         		& 10:07:00.59	& +67:32:47.2	& 142.90	&  42.65	& 14.327 	& 0.001		& 13.683	& 0.002		& 13.832		& 0.002			& 1			& 0			& 1	& 2.29	& 0.04	& 20.00	&      35	\\
 SW Sex                      	& PG 1012-029         		& 10:15:09.39	& -03:08:32.9	& 245.52	&  41.68	& 15.071 	& 0.002		& 14.952	& 0.004		& 14.815		& 0.003			& 1			& 0			& 1	& 2.53	& 0.10	& 19.62	&      25	\\
 GG Leo                      	& RX J1015.5+0904     		& 10:15:34.67	& +09:04:42.5	& 231.59	&  49.05	& 18.949 	& 0.014		& 18.936	& 0.047		& 19.122		& 0.032			& 0			& 0			& 1	& 1.65	& 0.07	& 20.20	&     -15	\\
 CP Dra                      	&                     		& 10:15:39.82	& +73:26:05.2	& 136.34	&  39.38	& 15.587 	& 0.002		& 15.628	& 0.005		& 15.796		& 0.004			& 0			& 0			& 0	& 1.44	& 0.07	& 20.58	&     -15	\\
 CI UMa                      	& SVS 1755            		& 10:18:12.96	& +71:55:43.6	& 137.59	&  40.55	& 18.490 	& 0.010		& 18.247	& 0.023		& 18.036		& 0.014			& 1			& 0			& 0	& 1.20	& 0.12	& 20.80	&      45	\\
 KS UMa                      	& RX J1020+5304       		& 10:20:26.52	& +53:04:33.2	& 159.55	&  51.93	& 16.775 	& 0.004		& 16.539	& 0.008		& 15.953		& 0.004			& 1			& 0			& 0	& 0.75	& 0.09	& 20.86	&     110	\\
 FIRST J102347+003841        	&                     		& 10:23:47.70	& +00:38:41.2	& 243.49	&  45.78	& 17.315 	& 0.005		& 16.856	& 0.010		& 17.315		& 0.009			& 0			& 0			& 1	& 1.83	& 0.07	& 20.41	&     -10	\\
 U Leo                       	& BD +14 2239         		& 10:24:03.92	& +14:00:25.3	& 226.34	&  53.27	& 17.007 	& 0.004		& 16.696	& 0.009		& 17.063		& 0.008			& 0			& 0			& 0	& 1.08	& 0.10	& 20.53	&     -10	\\
 DW UMa                      	& PG 1030+590         		& 10:33:52.86	& +58:46:54.8	& 150.28	&  50.41	& 15.211 	& 0.002		& 15.116	& 0.004		& 14.421		& 0.002			& 1			& 0			& 0	& 0.70	& 0.07	& 20.85	&     110	\\
 DO Leo                      	& PG 1038+155         		& 10:40:51.23	& +15:11:34.0	& 227.71	&  57.43	& 17.591 	& 0.006		& 16.947	& 0.010		& 17.136		& 0.009			& 1			& 0			& 0	& 1.32	& 0.04	& 20.64	&      30	\\
 IY UMa                      	& Tmz V85             		& 10:43:56.72	& +58:07:31.9	& 149.76	&  51.84	& 17.638 	& 0.006		& 17.299	& 0.012		& 17.046		& 0.008			& 0			& 2			& 1	& 2.31	& 0.12	& 20.11	&      65	\\
 CW 1045+525                 	&                     		& 10:48:18.01	& +52:18:30.0	& 156.89	&  55.92	& 14.608 	& 0.001		& 14.000	& 0.002		& 14.164		& 0.002			& 1			& 0			& 0	& 1.50	& 0.04	& 20.63	&      30	\\
 SX LMi                      	& CBS-31/Ton 45       		& 10:54:30.44	& +30:06:10.4	& 199.34	&  64.24	& 16.146 	& 0.003		& 15.917	& 0.006		& 15.596		& 0.004			& 1			& 0			& 0	& 1.36	& 0.05	& 20.42	&      60	\\
 CY UMa                      	& SVS 2198            		& 10:56:57.03	& +49:41:18.3	& 159.37	&  58.55	& 17.336 	& 0.005		& 16.995	& 0.011		& 16.534		& 0.006			& 1			& 0			& 0	& 1.39	& 0.05	& 20.42	&     100	\\
 AN UMa                      	& PG 1101+453         		& 11:04:25.62	& +45:03:14.5	& 165.83	&  62.15	& 16.251 	& 0.003		& 16.103	& 0.006		& 15.908		& 0.004			& 1			& 0			& 1	& 1.97	& 0.05	& 20.33	&      35	\\
 ST LMi                      	& CW 1103+254         		& 11:05:39.79	& +25:06:28.7	& 211.80	&  66.21	& 17.523 	& 0.005		& 16.681	& 0.009		& 17.343		& 0.010			& 0			& 0			& 0	& 1.30	& 0.07	& 20.49	&      -5	\\
 AR UMa                      	& 1ES 1113+432        		& 11:15:44.56	& +42:58:22.6	& 167.45	&  64.97	& 16.468 	& 0.003		& 15.962	& 0.006		& 16.269		& 0.005			& 0			& 0			& 0	& 0.75	& 0.06	& 20.77	&       5	\\
 DP Leo                      	& 1E 1114+182         		& 11:17:15.93	& +17:57:41.8	& 230.90	&  66.46	& 18.082 	& 0.007		& 17.915	& 0.018		& 17.723		& 0.012			& 1			& 0			& 1	& 2.06	& 0.02	& 20.11	&      35	\\
 TT Crt                      	& FSV 1132-11         		& 11:34:47.17	& -11:45:30.2	& 274.88	&  46.90	& 15.729 	& 0.002		& 15.156	& 0.004		& 15.379		& 0.003			& 1			& 0			& 1	& 2.78	& 0.11	& 19.48	&      20	\\
 RZ Leo                      	&                     		& 11:37:22.24	& +01:48:58.9	& 264.77	&  59.09	& 18.715 	& 0.011		& 18.013	& 0.020		& 17.932		& 0.014			& 1			& 0			& 1	& 2.47	& 0.03	& 19.77	&      80	\\
 T Leo                       	& BD +4 2506a         		& 11:38:26.81	& +03:22:07.5	& 263.49	&  60.52	& 16.048 	& 0.002		& 15.676	& 0.005		& 14.993		& 0.003			& 1			& 0			& 1	& 2.69	& 0.20	& 19.42	&     155	\\
 DO Dra                      	& YY Dra?             		& 11:43:38.46	& +71:41:20.9	& 130.30	&  44.46	& 14.516 	& 0.001		& 13.872	& 0.002		& 13.485		& 0.001			& 1			& 0			& 0	& 1.37	& 0.07	& 20.53	&     130	\\
 TW Vir                      	& PG 1142-041         		& 11:45:21.18	& -04:26:05.7	& 273.60	&  54.63	& 15.763 	& 0.002		& 15.199	& 0.004		& 15.112		& 0.003			& 1			& 0			& 0	& 1.74	& 0.05	& 20.33	&      60	\\
 EU UMa                      	& RE 1149+28          		& 11:49:55.71	& +28:45:07.5	& 202.54	&  76.33	& 17.283 	& 0.004		& 17.140	& 0.011		& 16.477		& 0.006			& 1			& 0			& 1	& 2.49	& 0.05	& 19.87	&     110	\\
 BC UMa                      	& GR 102              		& 11:52:15.84	& +49:14:41.9	& 146.27	&  65.12	& 18.157 	& 0.007		& 17.971	& 0.019		& 17.570		& 0.011			& 1			& 0			& 1	& 2.04	& 0.11	& 20.01	&      70	\\
 IR Com                      	& S 10932             		& 12:39:32.05	& +21:08:06.3	& 277.93	&  83.42	& 18.618 	& 0.010		& 17.809	& 0.018		& 18.534		& 0.020			& 0			& 0			& 0	& 1.08	& 0.04	& 20.55	&     -10	\\
 EV UMa                      	& RX J1307+53         		& 13:07:53.82	& +53:51:31.0	& 117.57	&  63.10	& 16.630 	& 0.003		& 16.759	& 0.009		& 16.495		& 0.006			& 1			& 0			& 0	& 1.07	& 0.06	& 20.65	&      20	\\
 HV Vir                      	& NSV 6201            		& 13:21:03.18	& +01:53:29.0	& 319.88	&  63.78	& 19.058 	& 0.014		& 19.013	& 0.040		& 18.938		& 0.025			& 0			& 2			& 0	& 1.18	& 0.07	& 20.74	&      10	\\
 HS Vir                      	& PG 1341-079         		& 13:43:38.44	& -08:14:03.6	& 324.44	&  52.44	& 15.206 	& 0.002		& 14.980	& 0.003		& 14.839		& 0.003			& 1			& 0			& 0	& 1.17	& 0.12	& 20.74	&      35	\\
 OU Vir                      	& 1432-0033           		& 14:35:00.23	& -00:46:06.2	& 348.90	&  52.60	& 16.556 	& 0.003		& 16.445	& 0.007		& 16.013		& 0.005			& 1			& 0			& 1	& 1.84	& 0.06	& 20.11	&      65	\\
 KV Dra                      	& RX J1450.5+6403     		& 14:50:38.34	& +64:03:28.8	& 103.81	&  48.41	& 17.086 	& 0.004		& 16.766	& 0.009		& 16.439		& 0.006			& 1			& 0			& 0	& 1.80	& 0.03	& 20.25	&      70	\\
 TT Boo                      	& HV 3681             		& 14:57:44.79	& +40:43:40.9	&  68.74	&  60.70	& 13.688 	& 0.001		& 13.722	& 0.002		& 13.740		& 0.002			& 0			& 2			& 0	& 0.74	& 0.07	& 20.99	&      -5	\\
 NY Ser                      	& NSV 6990            		& 15:13:02.33	& +23:15:08.5	&  33.99	&  57.84	& 15.311 	& 0.002		& 15.269	& 0.004		& 15.341		& 0.003			& 0			& 0			& 0	& 0.82	& 0.04	& 20.82	&      -5	\\
 ES Dra                      	& PG 1524+622         		& 15:25:31.80	& +62:00:59.8	&  97.64	&  46.83	& 15.019 	& 0.001		& 14.823	& 0.003		& 14.905		& 0.003			& 1			& 0			& 0	& 0.89	& 0.17	& 20.72	&       5	\\
 QW Ser                      	& Tmz V46             		& 15:26:13.99	& +08:18:02.5	&  13.06	&  48.86	& 17.603 	& 0.005		& 17.140	& 0.012		& 16.941		& 0.008			& 1			& 0			& 0	& 1.72	& 0.04	& 20.08	&      70	\\
 ASAS 153616-0839.1          	&                     		& 15:36:16.00	& -08:39:07.7	& 356.94	&  36.40	& 18.096 	& 0.007		& 17.852	& 0.018		& 16.993		& 0.008			& 1			& 0			& 0	& 0.80	& 0.11	& 20.76	&     175	\\
 LX Ser                      	& Stepanian's star    		& 15:38:00.11	& +18:52:03.5	&  29.78	&  50.97	& 14.762 	& 0.001		& 14.431	& 0.003		& 14.162		& 0.002			& 1			& 0			& 0	& 0.70	& 0.06	& 21.02	&      65	\\
 CT Ser                      	& N Ser 1948 (r)      		& 15:45:39.09	& +14:22:31.5	&  24.48	&  47.56	& 16.434 	& 0.003		& 16.351	& 0.007		& 16.360		& 0.006			& 1			& 0			& 0	& 0.83	& 0.04	& 20.96	&       5	\\
 SS UMi                      	& PG 1551+719         		& 15:51:22.37	& +71:45:12.1	& 106.37	&  39.06	& 16.001 	& 0.002		& 15.901	& 0.006		& 15.741		& 0.004			& 1			& 0			& 0	& 0.61	& 0.14	& 20.72	&      25	\\
 MR Ser                      	& PG 1550+191         		& 15:52:47.16	& +18:56:29.2	&  31.72	&  47.71	& 17.129 	& 0.004		& 16.468	& 0.008		& 17.003		& 0.008			& 0			& 0			& 1	& 1.65	& 0.04	& 20.39	&      -5	\\
 RX J1554.2+2721             	&                     		& 15:54:12.35	& +27:21:52.7	&  44.16	&  49.61	& 17.103 	& 0.004		& 16.243	& 0.007		& 16.418		& 0.006			& 1			& 0			& 1	& 1.66	& 0.08	& 20.21	&      55	\\
 QZ Ser                      	& HadV4               		& 15:56:54.56	& +21:07:19.9	&  35.24	&  47.49	& 17.192 	& 0.004		& 16.576	& 0.008		& 16.731		& 0.007			& 1			& 0			& 0	& 1.74	& 0.05	& 20.17	&      35	\\
 VW CrB                      	& Antipin V21         		& 16:00:03.71	& +33:11:14.2	&  53.29	&  49.12	& 16.652 	& 0.003		& 16.392	& 0.007		& 16.633		& 0.006			& 0			& 0			& 0	& 0.74	& 0.06	& 20.85	&      -5	\\
 1RXS J161008+035222         	&                     		& 16:10:07.55	& +03:52:33.1	&  15.81	&  37.26	& 17.596 	& 0.005		& 16.262	& 0.007		& 16.899		& 0.007			& 1			& 0			& 0	& 1.03	& 0.05	& 20.74	&      40	\\
 X Ser                       	& HV 3137             		& 16:19:17.70	& -02:29:29.3	&  10.84	&  31.87	& 17.050 	& 0.004		& 16.544	& 0.008		& 16.648		& 0.006			& 1			& 0			& 0	& 0.93	& 0.09	& 20.62	&      30	\\
 V589 Her                    	& S 10296             		& 16:22:07.19	& +19:22:36.8	&  35.56	&  41.33	& 17.610 	& 0.006		& 17.127	& 0.011		& 16.899		& 0.008			& 1			& 0			& 1	& 2.13	& 0.03	& 19.98	&      75	\\
 V844 Her                    	& Antipin V43         		& 16:25:01.76	& +39:09:26.6	&  62.36	&  44.38	& 17.656 	& 0.005		& 17.486	& 0.014		& 16.730		& 0.007			& 1			& 0			& 0	& 0.66	& 0.06	& 20.82	&     135	\\
 V699 Oph                    	& HV 10577            		& 16:25:14.79	& -04:40:25.7	&   9.73	&  29.41	& 15.915 	& 0.002		& 15.173	& 0.004		& 15.699		& 0.004			& 0			& 0			& 0	& 0.82	& 0.08	& 20.78	&       0	\\
 RW UMi                      	& SVS 1359            		& 16:47:54.84	& +77:02:12.2	& 109.64	&  33.15	& 18.677 	& 0.010		& 18.531	& 0.031		& 18.454		& 0.018			& 1			& 0			& 0	& 0.74	& 0.15	& 20.76	&      20	\\
 V841 Oph                    	& BD -12 4633         		& 16:59:30.38	& -12:53:26.8	&   7.62	&  17.78	& 13.344 	& 0.001		& 12.937	& 0.001		& 13.203		& 0.001			& 0			& 2			& 0	& 1.09	& 0.07	& 20.53	&       5	\\
\hline                                                                                                                                                                 
\multicolumn{10}{l}{} \\                                                                                                                                               
\multicolumn{10}{l}{\textbf{Plane}} \\                                                                                                                     
\hline                                                                                                                                                                 
    HT Cas                   	&            S 3343   		& 01:10:13.17	& +60:04:35.6	& 125.27	& -2.71 	& 15.936 	& 0.004		& 15.455	& 0.005		& 15.056		& 0.003			& 1			& 0			& 0	& 1.24	& 0.06	& 20.06	&        110	\\
     V Per                   	&       BD +56 406a   		& 02:01:53.93	& +56:44:03.5	& 132.52	& -4.82 	& 17.983 	& 0.014		& 17.746	& 0.018		& 17.378		& 0.011			& 1			& 0			& 0	& 0.85	& 0.06	& 20.55	&         70	\\
    UV Per                   	&        AN 87.1911   		& 02:10:08.30	& +57:11:21.0	& 133.47	& -4.06 	& 18.195 	& 0.015		& 17.763	& 0.018		& 17.110		& 0.009			& 1			& 0			& 0	& 0.85	& 0.06	& 20.67	&        160	\\
    TZ Per                   	&        AN 28.1912   		& 02:13:50.95	& +58:22:52.4	& 133.58	& -2.78 	& 12.673 	& 0.001		& 12.465	& 0.001		& 12.586		& 0.001			& 0			& 0			& 0	& 1.02	& 0.08	& 20.51	&          5	\\
     T Aur                   	&       BD +30 923a   		& 05:31:59.12	& +30:26:45.4	& 177.14	& -1.70 	& 14.910 	& 0.003		& 14.573	& 0.003		& 14.603		& 0.003			& 1			& 0			& 0	& 1.25	& 0.04	& 20.38	&         25	\\
    FS Aur                   	&            S 3946   		& 05:47:48.37	& +28:35:11.2	& 180.55	&  0.23 	& 15.797 	& 0.004		& 15.350	& 0.004		& 15.241		& 0.004			& 1			& 0			& 0	& 0.99	& 0.10	& 20.52	&         55	\\
    CW Mon                   	&        AN 61.1936   		& 06:36:54.59	& +00:02:17.3	& 211.24	& -3.21 	& 18.339 	& 0.094		& 17.094	& 0.049		& 17.180		& 0.039			& 0			& 0			& 1	& 1.17	& 0.05	& 18.45	&        130	\\
  V603 Aql                   	&         HD 174107   		& 18:48:54.64	& +00:35:03.0	&  33.16	&  0.83 	& 12.425 	& 0.001		& 11.995	& 0.001		& 11.835		& 0.001			& 0			& 2			& 0	& 0.88	& 0.08	& 20.67	&         60	\\   
    CI Aql                   	&        AN 23.1925   		& 18:52:03.57	& -01:28:39.0	&  31.69	& -0.81 	& 15.562 	& 0.002		& 14.812	& 0.003		& 15.438		& 0.004			& 0			& 0			& 0	& 1.08	& 0.07	& 20.54	&         -5	\\
  V446 Her                   	&                     		& 18:57:21.61	& +13:14:29.7	&  45.41	&  4.71 	& 16.759 	& 0.005		& 16.242	& 0.008		& 16.443		& 0.007			& 1			& 0			& 0	& 1.47	& 0.08	& 19.73	&         20	\\
 V1425 Aql                   	&                     		& 19:05:26.66	& -01:42:03.3	&  33.01	& -3.89 	& 18.494 	& 0.011		& 18.693	& 0.042		& 15.731		& 0.004			& 0			& 2			& 0	& 1.30	& 0.10	& 20.29	&     456955\tablenotemark{e}	\\
    EM Cyg                   	&       AN 185.1928   		& 19:38:40.13	& +30:30:28.6	&  65.19	&  4.28 	& 13.129 	& 0.001		& 12.697	& 0.001		& 12.881		& 0.001			& 1			& 0			& 0	& 1.84	& 0.09	& 19.57	&         15	\\
 RX J1951.7+3716             	&                     		& 19:51:47.50	& +37:16:48.1	&  72.44	&  5.27 	& 14.961 	& 0.002		& 14.393	& 0.003		& 14.353		& 0.003			& 1			& 0			& 1	& 1.77	& 0.07	& 18.91	&         55	\\   
    EY Cyg                   	&       AN 200.1928   		& 19:54:36.77	& +32:21:55.5	&  68.50	&  2.26 	& 14.125 	& 0.001		& 13.760	& 0.002		& 13.878		& 0.002			& 1			& 0			& 1	& 1.48	& 0.05	& 19.19	&         15	\\
 V2306 Cyg                   	& 1WGA J1958.2+3232   		& 19:58:14.47	& +32:32:42.2	&  69.05	&  1.70 	& 15.479 	& 0.002		& 15.240	& 0.004		& 15.092		& 0.003			& 1			& 0			& 0	& 0.91	& 0.06	& 20.38	&         35	\\
  V550 Cyg                   	&            S 3847   		& 20:05:04.95	& +32:21:22.9	&  69.66	&  0.38 	& 16.621 	& 0.003		& 15.922	& 0.006		& 16.544		& 0.006			& 0			& 0			& 0	& 1.12	& 0.07	& 20.63	&        -10	\\
 V503 Cyg                    	&            S 4524   		& 20:27:17.39	& +43:41:22.5	&  81.50	&  3.07 	& 17.252 	& 0.013		& 16.842	& 0.014		& 16.404		& 0.011			& 1			& 0			& 0	& 1.01	& 0.07	& 19.82	&        105	\\
 V751 Cyg                    	&          SVS 1202   		& 20:52:12.80	& +44:19:26.2	&  84.74	& -0.09 	& 13.879 	& 0.002		& 13.653	& 0.002		& 13.712		& 0.002			& 0			& 0			& 0	& 1.00	& 0.06	& 20.59	&         10	\\
 V2275 Cyg                   	& N Cyg 2001, No. 2   		& 21:03:01.98	& +48:45:53.5	&  89.32	&  1.39 	& 17.709 	& 0.011		& 17.017	& 0.011		& 17.312		& 0.010			& 1			& 0			& 1	& 3.04	& 0.08	& 19.24	&         25	\\                                                                                                                                                                            
\footnotetext[1]{Possible entries in this column; 0 = not detected as a H$\alpha$ emitter, 1 = detected as a H$\alpha$ emitter}                                             
\footnotetext[2]{Possible entries in this column; 0 = no flag, 2 = incorrect object classification or object too bright, 3 = magnitudes missing, 5 = situated on bad pixels}
\footnotetext[3]{Possible entries in this column; 0 = no flag, 1 = bad quality field that is, seeing $>$ 2 or ellipticity $<$ 0.2 or limiting magnitude $<$ 19.5 in any of the three bands}
\footnotetext[4]{Values are estimates obtained from IPHAS photometry.}
\footnotetext[5]{This value is too large to be physically real.}

\end{longtable}      
\end{landscape}   
\normalsize
\twocolumn
\subsection{Interpreting H$\alpha$ Excess and Estimating Equivalent Widths}
\label{ews}

It is tempting to treat the H$\alpha$ excess values calculated
from Equation~1 as direct tracers of emission line strength
(tracers of line EW). However, this interpretation is not always
safe. In particular, it is important to keep in mind that the
calculated excess values have no fixed reference point, that is each
excess is calculated relative to the  
local fit to the upper boundary of the stellar locus in the
appropriate magnitude bin of a given field. Thus any direct comparison of
H$\alpha$ excess values between different objects implicitly assumes that
each object actually belonged to the stellar population that was
described by the fit. In fields containing a split stellar locus
(and hence more than one population), there is no guarantee that this
is the case. 

It is worth stressing that this problem is not simply due
to the particular definition of ``H$\alpha$ excess'' we have
adopted. Instead, it reflects the inherent ambiguity in assigning
individual objects to particular stellar populations on the basis of
limited photometric information. More specifically, since different
stellar populations can have different ``baseline''
$(r^\prime-$H$\alpha)$ colours at given $(r^\prime-i^\prime)$, it is
impossible to define a physically meaningful H$\alpha$ {\em excess}
for a given source without specifying the population to which it
belongs.

This all sounds rather pessimistic. However, we can also turn this
argument around: {\em if we are willing to adopt a particular 
spectral energy distribution (SED) for a given H$\alpha$ excess 
source, a corresponding H$\alpha$ EW estimate can be obtained
immediately from the photometric data.} If this SED represents the
objects'continuum distribution better than the fit to the upper
stellar locus, the resulting EW estimate should be a more meaningful
quantity than the straight H$\alpha$ excess.

\begin{figure*}
\includegraphics[angle=270,width=\textwidth]{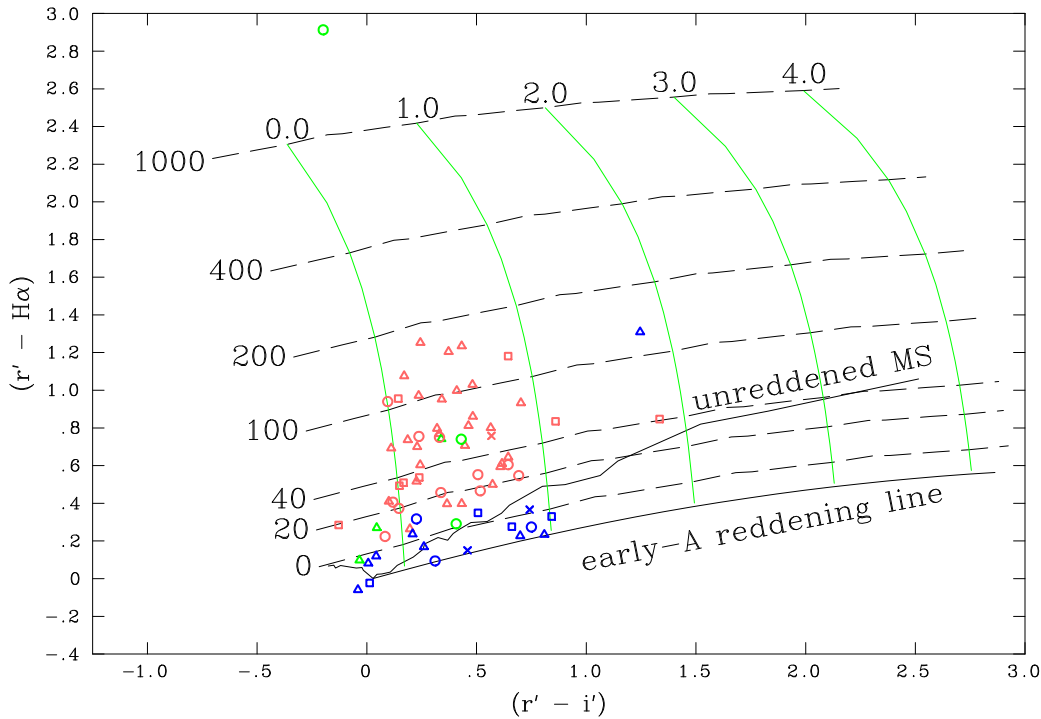}
\caption{\label{eq_width} Colour-colour plot showing how H$\alpha$
EW relates to H$\alpha$ excess and broad-band colour.    Triangles
represent dwarf novae; circles represent novae, novalike variables and
recurrent novae; squares represent magnetic systems and crosses
represent non-classified systems.  The symbols are plotted in red if
the system is a H$\alpha$ emitter; blue if it is a non-emitter; and
green if it is a system which did not pass our initial selection cuts.
The jagged solid curve is the unreddened main sequence (MS) line, the smooth solid curve running mainly horizontally is the
early-A reddening line (that is the reddening track of an early A
star). All other lines correspond to power law SEDs with 
$F_\lambda
\propto \lambda^{-2.3}$, as appropriate for an optically thick
accretion disc (see \citealt{2005MNRAS.362..753D}). The dashed lines are
lines of constant H$\alpha$\ EW superposed on this SED (the EW values
are given in Angstroms at the far left of each line). The solid green
vertical lines represent H$\alpha$ emission curves of growth for this
SED at fixed reddening (the corresponding E(B - V) values are
shown at the top of each line).}
\end{figure*}

Fig.\,\ref{eq_width} illustrates how this idea can be implemented for
CVs. Basically, the location of a given source in the colour-colour
diagram is determined by its H$\alpha$\ EW and the shape of its
continuum SED. The latter, in  
turn, depends on the intrinsic SED shape and the reddening towards the  
source. For CVs, it may be reasonable to adopt an intrinsic SED of $F_\lambda
\propto \lambda^{-2.3}$\ as typical. This is appropriate for an optically
thick accretion disc and is probably acceptable even for
WD-dominated short-period DNe. We can then add H$\alpha$ lines of
varying EWs to this SED and calculate the reddening tracks of the
resulting model spectra through colour-colour space. As shown in
Fig.\,\ref{eq_width}, this procedure yields a mapping between the observed
colours, on the one hand, and the corresponding E$_{B-V}$ values and
H$\alpha$ EWs, on the other. 

In principle, it should be possible to place observed points directly
onto this mapped colour-colour plot and read off the resulting
E$_{B-V}$ values and H$\alpha$ EWs. However, in practice, one 
additional correction is necessary at present. This arises because our
synthetic photometry is Vega-based, but the standard stars used for
the photometric calibration of the IPHAS data are mostly later type
stars. As noted by \citet{2005MNRAS.362..753D}, Vega is an A0V star
with deep H$\alpha$ absorption, and this causes a systematic offset
between observed and synthetic $r^\prime$ - H$\alpha$ colours.  The offset 
required for a given field is between 0.25 and -0,05 mag (in the sense that the data points must be shifted
upwards). We thus account for this by applying an average shift of 0.15 mag 
to the observed $r^\prime$ - H$\alpha$ colours of all CVs before
placing them on Fig.\,\ref{eq_width}. 

It is also worth noting that the IPHAS data currently still lack a
global (field-to-field) photometric calibration. This again
particularly affects the H$\alpha$ photometry and causes a scatter of
$\approx 0.1$ mag in the $r^\prime$ - H$\alpha$ colours between different
fields. We currently make no attempt to correct for this. Note that
both systematic and field-to-field offsets will be automatically
corrected in the final IPHAS data release.

Fig.\,\ref{eq_width} shows the corrected colours of our CVs on the
mapped colour-colour diagram. The resulting H$\alpha$ EW estimates are
given in Table~\ref{tab1_b}, and we will adopt these values throughout
the rest of this paper. We stress that these may be subject to
considerable error, both because of field-to-field scatter and because
the intrinsic SED shape we have assumed may not be appropriate to all
systems. \footnote{On a more positive note, Figure~6 in
\citet{2005MNRAS.362..753D} shows that, for {\em smooth} SEDs, most of
the sensitivity to the assumed SED shape is in the reddening
estimates, whereas H$\alpha$ EWs are much less affected.}

\section{Results}
\label{res_s}

Table~\ref{tab1_a} summarizes the key properties of all the CVs in the
two samples. The magnitude range data were taken from the
Downes \textit{et al}
catalogue\footnote{http://archive.stsci.edu/prepds/cvcat/index.html}
\citep{2001PASP..113..764D}, orbital periods are from the Ritter 
\& Kolb catalogue, and the CV type information comes from the
web-based CV catalogue CVCat\footnote{http://cvcat.net} 
\citep{2003A&A...404.1159K}. Our 
new data are listed Table~\ref{tab1_b}, which also includes
information on whether a given CV was selected as an H$\alpha$
emitter.

\subsection{Recovery Rates of CVs as H$\alpha$ Emitters}

\begin{figure*}
\includegraphics[angle=0,width=\textwidth]{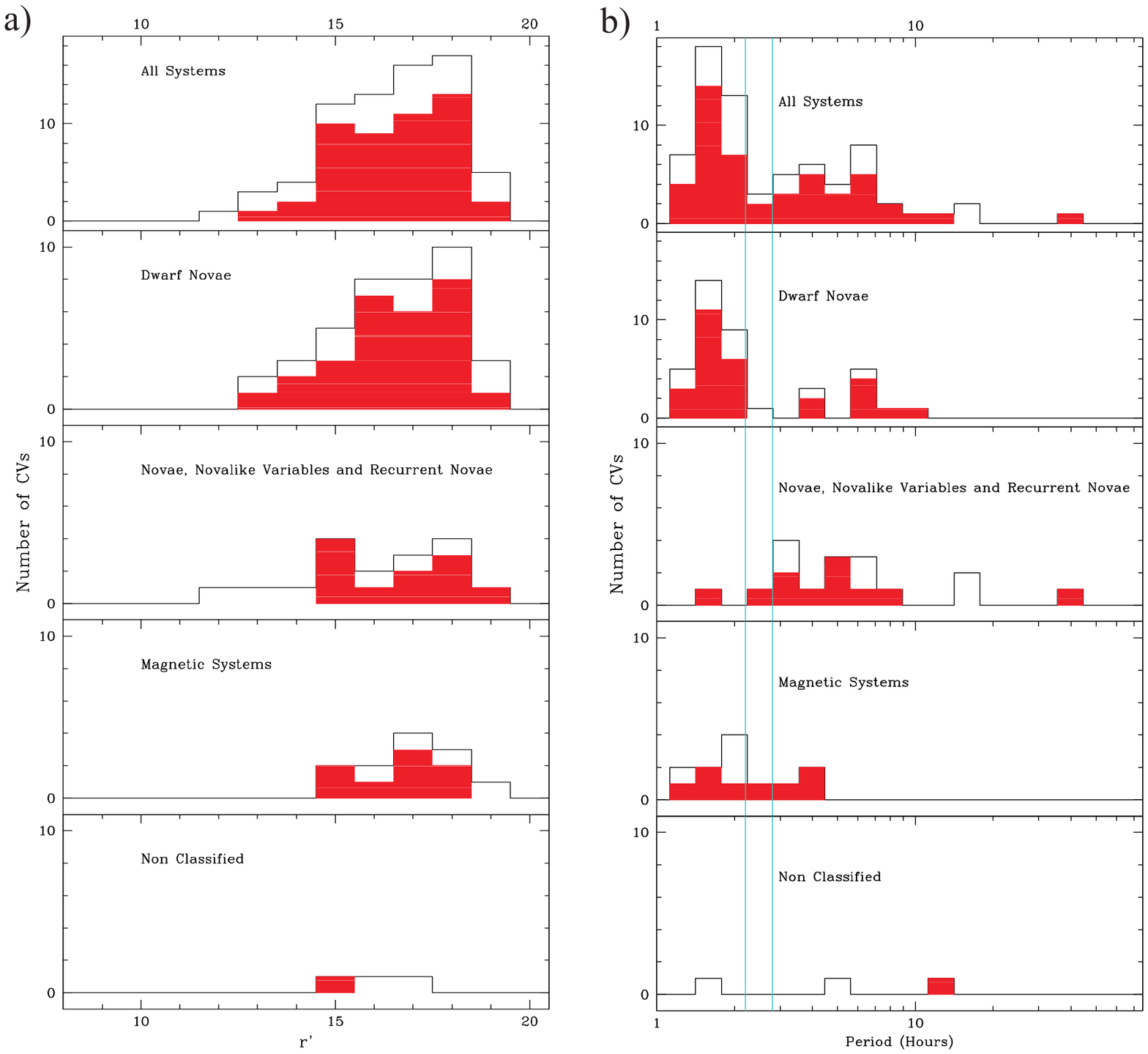}
\caption{\label{mag_hist} a) The magnitude distribution of the CVs
b) the period distribution of the CVs. The orbital period gap is shown
between the cyan lines.  The plots show the
distribution of emitters in red overlaid on top of the total
distribution of CVs.}
\end{figure*}

Our recovery rates are summarized in Tables~\ref{tab2}-\ref{mag_stats}
and illustrated in Figs~\ref{mag_hist}-\ref{gal_non}. In the following
subsections, we discuss the resulting completeness 
estimates and their dependence on key CV properties (orbital period,
CV type, apparent magnitude, galactic latitude). We also
explore the causes that prevent some CVs from being selected as H$\alpha$
emitters. 

\begin{table}
\caption{\label{tab2} The distribution of H$\alpha$ emitters amongst
the CV samples. The numbers are given as a ratio of CVs
selected as H$\alpha$ emitters to the total number of CVs in the
sample. The numbers in parenthesis show the corresponding percentages.}
\begin{tabular}{lll}
\hline Sample & All CVs & CVs Passing \\
&& initial cut \\
\hline
Off-Plane 	& 36/52 (69.2)	 & 36/48 (75.0) \\
In-Plane 	& 12/19 (63.2)	 & 12/17 (70.6) \\
Total 		& 48/71 (67.6)	 & 48/65 (73.8) \\ \hline
\end{tabular}
\end{table}

\subsubsection{Overall Completeness}

Table~\ref{tab2} gives the recovery rates (that is CVs selected as
H$\alpha$\ emitters) for both in-plane and off-plane samples and also
the combined recovery rates. We find a global recovery rate of
$\simeq$ 68\% [48/71] for all CVs and $\simeq$ 74\% [48/65] if we only
consider  those CVs that passed our initial selection cuts defined in
section~\,\ref{initial_selection}. 

Both of these numbers are important. The first provides an
``end-to-end'' estimate of the survey completeness  for CVs, that is what
fraction of CVs brighter than $r^\prime =
19.5$ should we expect to discover with IPHAS (or similar surveys) as
emission line objects. This estimate allows for incompleteness due to
crowding/blending, objects falling on bad pixels, astrometric
calibration problems, etc. By contrast, the second completeness
estimate (considering only objects that pass our initial selection
cuts) is an estimate of the fraction of CVs that are {\em in
principle} discoverable by IPHAS, that is the fraction of CVs that
exhibit photometrically detectable H$\alpha$ excesses. Note that the number of CVs detected as H$\alpha$ emitters is the same in both
cases, since a CV which does not pass the initial selection cuts can
never be selected as an H$\alpha$ emitter.

It is interesting to ask what constitutes a ``detectable''
H$\alpha$ excess in this context. A useful perspective on this
question can be obtained from Fig.\,\ref{eq_width}, which 
shows that the unreddened main sequence broadly separates sources 
detected as H$\alpha$\ emitters from sources not detected as
H$\alpha$\ excess objects. This reflects the fact that our selection
procedure uses the upper boundary of the main stellar locus as the
zero point in searching for ``significant'' H$\alpha$\ excess outliers
and confirms that this boundary usually tracks the unreddened main
sequence. 

It is worth noting that this effectively constitutes an observational
selection bias. For a given SED, reddened sources must have larger
H$\alpha$\ EWs than unreddened sources in order to be selected as
H$\alpha$\ excess objects. Most CVs are blue objects observed through
relatively little extinction, so the sense of this selection bias
works in their favour. In essence, CVs occupy a region in the
colour-colour plots where even a small H$\alpha$ EW stands out
prominently from the stellar locus. The high recovery rate we find is
partly due to this fortunate situation.

\begin{table}
\caption{\label{per_stats} A summary of the results showing the
distribution amongst the long and short period CVs. Numbers are given as a ratio of CVs selected as H$\alpha$
emitters to the total number of CVs in the sample, the percentages are
given parenthesis.}
\begin{tabular}{llll}
\hline & Sample & All CVs & CVs passing \\
 &&& initial cut \\
\hline
Short & Off-Plane & 21/34 (61.8) & 21/31 (67.7) \\
Period & In-Plane & 4/5 (80.0) & 4/5 (80.0) \\
$<$ 2.5 hr & Total & 25/39 (64.1) & 25/36 (69.4) \\
\hline
Long & Off-Plane & 15/18 (83.3) & 15/17 (88.2) \\
Period & In-Plane & 8/14 (57.1) & 8/12 (66.7) \\
$>$ 2.5 hr & Total & 23/32 (71.9) & 23/29 (79.3) \\ \hline
\end{tabular}
\end{table}

\subsubsection{Dependence on Orbital Period}
\label{orb_per_s}

In order to assess the dependence of our completeness on orbital
period, we first compare the recovery statistics of short-period CVs
(defined as having orbital periods below the 2.2 - 2,8 hour period gap) to
those of long-period CVs (with periods above the gap). These recovery
rates for both types of CVs are listed in  Table~\ref{per_stats}. 

Somewhat surprisingly, the recovery rates for long- and short-period
systems are comparable, with both lying near 
65 per cent. Given that short-period CVs tend to be intrinsically faint and
that the H$\beta$\ EW and absolute magnitude for DNe are correlated (in
the sense that the faint systems have stronger lines; Patterson 1984), one
might have expected that short-period systems would be easier to
detect and thus have higher recovery rates. The fact that we do not
observe a marked difference in recovery rates is mainly a
testament to the high data quality of IPHAS, that is its ability to
separate even long-period objects with relatively weak emission lines
from the main stellar loci in any given field. This explanation is
supported by the fact that short-period DNe {\em do} exhibit the
largest H$\alpha$\ EWs (see section~\,\ref{excess_vs_porb} and
particularly Fig.\,\ref{excess_period}).

\begin{table}
\caption{\label{tab_type} Results showing the distribution of
H$\alpha$ emitters amongst the two samples of CVs.  The results in this
case are separated by CV
type. Numbers are given as a ratio of CVs selected as H$\alpha$
emitters to the total number of CVs in the sample, and the percentages
are shown in parenthesis.}
\begin{tabular}{llll}
\hline
CV type & Sample & All CVs & CVs passing \\
&&& initial cut \\
\hline
Dwarf			& Off-Plane 	& 22/30 (73.3) 	& 22/27 (81.5) \\
Novae 			& In-Plane 	& 6/9 (66.7) 	& 6/9 (66.7) \\
			& Total 	& 28/39 (71.8) 	& 28/36 (77.8) \\
\hline

Magnetic		& Off-Plane 	& 7/11 (63.6) 	& 7/11 (63.6) \\
Systems 		& In-Plane 	& 1/1 (100.0) 	& 1/1 (100.0) \\
			& Total 	& 8/12 (66.7) 	& 8/12 (66.7) \\
\hline

Nova, Recurrent		& Off-Plane 	& 7/9 (77.8) 	& 7/8 (87.5) \\
Nova and Novalike	& In-Plane 	& 4/8 (50.0) 	& 4/6 (66.7) \\
Variables		& Total 	& 11/17 (64.7) 	& 11/14 (78.6) \\
\hline

Non 			& Off-Plane 	& 0/2 (0.0) 	& 0/2 (0.0) \\
Classified 		& In-Plane 	& 1/1 (100.0) 	& 1/1 (100.0) \\
			& Total 	& 1/3 (33.3) 	& 1/3 (33.3) \\
\hline
\end{tabular}
\end{table}

\subsubsection{Dependence on CV Type}

In order to examine whether H$\alpha$\ selection favours particular CV
sub-types, each system in our samples is assigned one of four basic
types based on their classification in CVCat shown
in Table~\ref{tab1_a}: (i) dwarf novae; (ii) novae and nova-likes; (iii) magnetic CVs;
(iv) unclassified systems. Note that the second class includes both 
classical and recurrent novae, and the third class includes both
intermediate polars and the strongly magnetic polars. We refrain from using a finer classification scheme since
this would result in classes with extremely small numbers. It should also
be pointed out that some intermediate polars are known to undergo
dwarf nova outbursts. For the purposes of the work presented here, any
system which has been classified as a magnetic CV is put into this
category, irrespective of any similarities with weakly-magnetic systems.

Table~\ref{tab_type} shows the recovery statistics as a function of CV
types and Fig.\,\ref{mag_hist} shows how the recovery statistics depend on CV magnitude
and orbital period for the different CV types. The dwarf novae have the
overall highest recovery rate, which is not surprising since, in quiescence,
the accretion discs in these systems are at least partially optically
thin and are well known to produce strong emission lines (see \citealt{1995cvs..book.....W} and references therein).  The magnetic systems show the second highest recovery rate of all
systems. This is probably due to strong Balmer emission from their
accretion streams \citep{1990SSRv...54..195C, 1993MNRAS.265..605F}.

The novae and nova-likes have the lowest overall recovery rates, and
it is tempting to attribute this to their known observational and
physical characteristics. At first sight, this seems reasonable: both
types of high-state, weakly-magnetic CVs tend to exhibit only weak
emission lines when viewed at moderate inclinations (for example \citealt{1986ApJ...311..163S, 1987MNRAS.227...23W, 1996cvro.coll....3D}).
This is probably because these
systems contain optically thick discs that produce mainly absorption
line spectra (although often emission reversals are observed in the
cores of absorption line profiles).

However, a closer look at the statistics suggests that the true
explanation for the low recovery rates for these systems is more
prosaic. In the last column of Table~\ref{tab_type}, we show the
recovery rates that result if we restrict attention to systems that
have passed the initial selection cut. Here, the recovery fraction of
novae and nova-likes is the highest. These
numbers are completely consistent with the corresponding ones for DNe
and magnetic systems. Thus the reason for the lower overall recovery
statistics for novae and nova-likes is simply that relatively more of
these systems are misclassified as non-stellar in the IPHAS data.

It is worth asking why, despite the physical arguments
given above, the recovery rates for novae and nova-likes are
comparable to those of DNe. This question has actually already been
answered implicitly in section~\,\ref{orb_per_s}. There, we noted that
short-period CVs (which are dominated by DNe) do exhibit larger
H$\alpha$ excesses than long-period CVs (which contain most of the
novae and nova-likes), but that IPHAS is sensitive enough to still
identify the weaker H$\alpha$ emission often seen in the
latter.

Finally, the non-classified systems systems show a comparatively low
recovery rate in comparison to other systems. Given that there are only 3 non-classified 
CVs it is not prudent to draw any firm conclusions from the low recovery rate.

\begin{table}
\caption{\label{mag_stats} A summary of the results, in this case
showing how the recovery rate of H$\alpha$ emitters varies between the
four magnitude bins. The numbers are
given as a ratio of CVs selected as H$\alpha$ emitters to the total
number of CVs in the sample and the percentages are shown in
parenthesis}
\begin{tabular}{llll}
\hline
Magnitude  & Sample & All CVs & CVs passing \\
range ($r^\prime$) &&& initial cut \\
\hline
		& Off-Plane 	& 10/16 (62.5) 	& 10/14 (71.4) \\
 $<$ 16		& In-Plane 	& 7/11 (63.6) 	& 7/10 (70.0) \\
		& Total 	& 17/27 (63.0) 	& 17/24 (70.8) \\
\hline
		& Off-Plane 	& 14/19 (73.7) 	& 14/19 (73.7) \\
16 - 17.5 	& In-Plane 	& 2/3 (66.7) 	& 2/3 (66.7) \\
		& Total 	& 16/22 (72.7) 	& 16/22 (72.7) \\
\hline
		& Off-Plane 	& 10/12 (83.3) 	& 10/11 (90.9) \\
17.5 - 18.5 	& In-Plane 	& 3/5 (60.0) 	& 3/4 (75.0) \\
		& Total 	& 13/17 (76.5) 	& 13/15 (86.7) \\
\hline
		& Off-Plane 	& 2/5 (40.0) 	& 2/4 (50.0) \\
18.5 - 19.5 	& In-Plane 	& 0/0 (N.A.)  	& 0/0 (N.A.) \\
		& Total 	& 2/5 (40.0) 	& 2/4 (50.0) \\
\hline
\end{tabular}
\end{table}

\subsubsection{Dependence on CV Magnitude}
\label{faint_dependence} 

As photometric scatter increases towards fainter magnitudes,
completeness may be expected to decrease. Clearly, a faint CV will on
average need to display a greater H$\alpha$ excess in order to be
detected. Table~\ref{mag_stats} shows how the recovery rate of 
H$\alpha$ emitters depends on the magnitudes bins used in the
selection. It can be seen that for the Off-Plane sample the recovery rate
actually increases slightly as we go to fainter bins until we get to the
faintest bin, where it drops to 40 per cent (but note that there are
only 5 sources in this bin). The In-Plane sample also
hovers at a near constant 60 to 70 per cent for the three brighter
bins, and for this sample the faintest bin contains no objects. We
conclude that completeness for CVs in IPHAS remains steady and high up
to at least $r^\prime \simeq 18.5$. The small numbers of fainter CVs in
our samples prevent us from constraining the drop in completeness
beyond this point more precisely.

\subsection{Reasons for CVs Not Being Selected as H$\alpha$ Emitters}

There are several reasons that can prevent a CV from being selected as
an H$\alpha$ emitter: 

\begin{enumerate}
\item
The CV is not located significantly above the main stellar locus in
the colour-colour plots. In some cases, this may be due to weakness or
absence of an H$\alpha$\ emission line (as seen in erupting dwarf
novae and some nova-like variables). Furthermore, the equivalent 
width of the H$\alpha$ emission line can sometimes change
significantly during the orbital cycle, so some CVs may have 
been observed at phases where the emission is weak. Finally, at fainter
magnitudes, the signal-to-noise may simply become insufficient to
separate a CV from the main stellar locus.
\item
The CV is located below the locus in the colour-colour plots, that is
the CV shows H$\alpha$ absorption. This may again point towards an 
erupting dwarf nova or a nova-like variable.
\item
The fitting process failed, and the CV is not selected due to the
resulting selection cut being inappropriate. 
\item
The CV does not pass the initial selection cuts defined in
section~\,\ref{initial_selection} 
\end{enumerate}

In the following sections, we look in more detail at the various CV
sub-types and check if we can understand why particular CVs were not
recovered as emission line objects.

\subsubsection{Dwarf Novae}

The DNe are the most numerous class and generally show some of the
largest H$\alpha$ excesses. However, they also show the greatest range
in H$\alpha$ excesses, although this can be at least partly attributed
to observing some systems in quiescence and others in
outburst.  RZ LMi, CP Dra, TT Boo, NY Ser, and TZ Per are the DNe that
have been observed in outburst.  None of them have been detected as
H$\alpha$ emitters. Of the 6 DNe remaining that were not selected as H$\alpha$
emitters, IY UMa has a large H$\alpha$ excess but does not pass the initial selection cuts, for the reasons given in section~\,\ref{in-pl_sample}, and hence is not selected. CW Mon also has a large H$\alpha$ excess but is not selected
as described in section~\,\ref{excess_vs_porb}.  HV Vir misses our initial selection cuts and hence is not selected, however its EW shows that H$\alpha$ emission is present.  This leaves 3 systems which are not recovered as H$\alpha$ emitters: IR Com, VW CrB and V550 Cyg. These systems pass our initial selection cuts and do not have a field quality flag
associated with them.  It is likely that variations in EW due to orbital phase can account for the absence of observable H$\alpha$ emission in these systems, for example IR Com is known to be an eclipsing system and the IPHAS magnitudes are consistent with this system being observed in eclipse. 

The recovery rate for all the DNe in quiescence is 81 per
cent\footnote{The CVs HS Vir and OU Vir have been excluded from this
recovery rate calculation as the photometry suggests they are in an
intermediate state between quiescence and outburst.} which is 9 per
cent higher than the recovery rate for all the DNe including those in
outburst. 

\subsubsection{Magnetic CVs}

There are rather few magnetic systems in our samples (including only
one intermediate polar), but the majority show a detectable H$\alpha$
excess. Those which do not show emission are all AM Her systems
observed in the low state (GG Leo, ST LMi, AR UMa, and MR Ser). A 
reduction in H$\alpha$ emission line strength is a known
characteristic of low state magnetic CVs. It is possible that the
H$\alpha$ emission is not strong enough to be detected above Zeeman
absorption in these systems. 

\subsubsection{Novae, Recurrent Novae and Novalike Variables}

Many of the bluest CVs of this class show a large H$\alpha$ excess
indicating that the nova remnants are undergoing mass transfer. 
However, there is no simple link between detection of H$\alpha$
excess and time since eruption. For example, V Per and T Aur erupted
in 1891 and 1887, respectively, and both have been detected as
emitters. Conversely, the two oldest nova remnants -- V841 Oph (1848)
and U Leo (1855, though its nature as a nova is not certain) -- have not
been detected as H$\alpha$ excess objects. 

There is one nova with a very high H$\alpha$ excess that did not pass
the initial selection cut. This system is V1425 Aql. Spectroscopic
observations have shown strong H$\alpha$ emission both before and 
after outburst \citep{1997AJ....114.2671K}, which is consistent with
the large excess we observe.

\begin{figure*}
\includegraphics[angle=0,width=\textwidth]{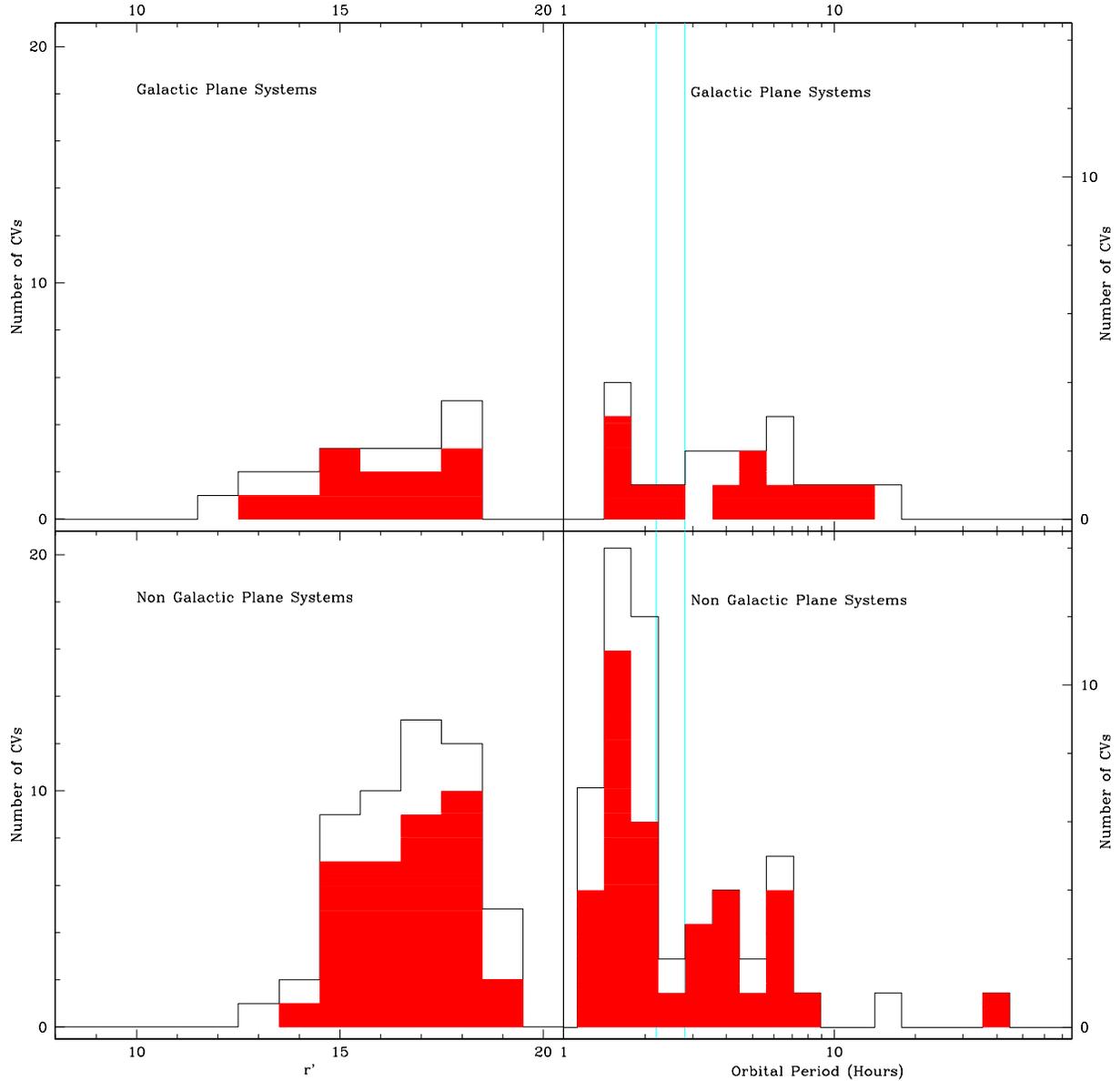}
\caption{\label{gal_non} The magnitude distribution and period
distribution of CVs showing those located within the galactic plane (upper
plots) and outside of the plane (lower plots). The plots show the
distribution of emitters in red overlaid on the total distribution
of CVs.    The orbital period gap is shown between
the cyan lines.}
\end{figure*}

\subsection{In-Plane vs Off-Plane CVs}
\label{gal_pla}

Given that IPHAS is a galactic plane survey, but that the majority of
the CVs analyzed here are Off-Plane sources, it is worth checking
whether crowding and extinction affects the recovery rate in galactic
plane fields. We have already provided separate recovery rates for
both samples in Tables~\ref{tab2}-\ref{mag_stats}, but in
Fig.\,\ref{gal_non} we now additionally present a visual comparison
between the recovery statistics of In-Plane and Off-Plane samples.

In general, we find that the recovery rates of the two samples are
similar. This suggests that crowding and extinction have at most a
small effect on completeness. However, despite the similar recovery
rates, there is nevertheless a clear difference in data between galactic 
plane fields and non-galactic plane fields. For example, when the
value of the rms scatter about the initial least square fits to the
stellar loci is compared between In-Plane and Off-Plane fields, the
scatter is clearly larger for the former. This is mainly due to
different populations seen at different distances through varying extinction
in galactic plane fields (though probably also to increased
photometric scatter due to higher crowding levels). Our
$\sigma$-clipping fits to the upper boundary of the main stellar
locus help partly to counter this effect. Without this technique, the
In-Plane recovery rates would be reduced.

\section{Correlations Between Physical and Photometric Properties}
\label{correl}

\subsection{Orbital Period vs H$\alpha$ EW}
\label{excess_vs_porb}

\begin{figure*}
\includegraphics[angle=0,width=\textwidth]{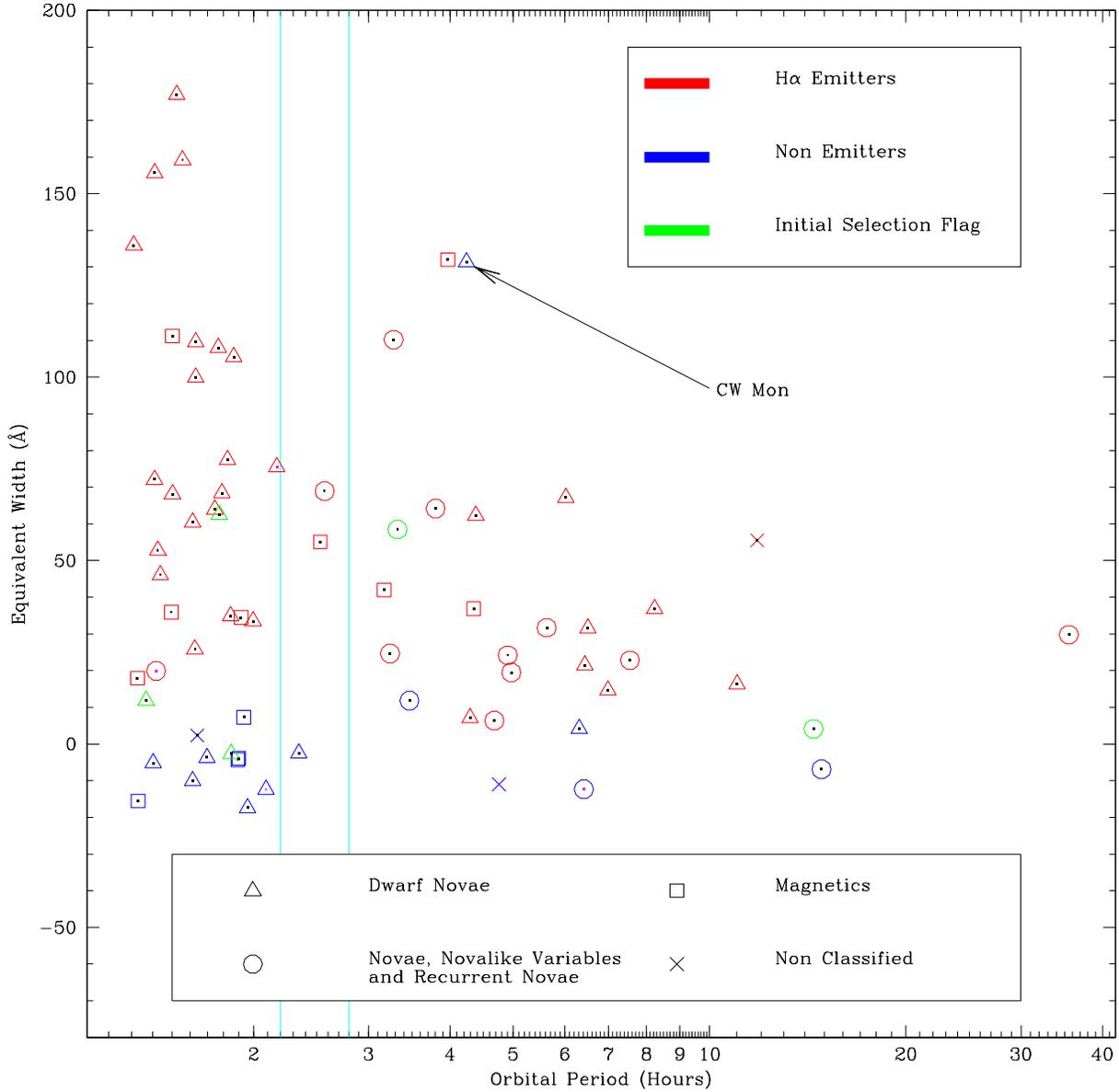}
\caption{\label{excess_period} A plot of EW versus the
orbital period of CVs in the Off-Plane and In-Plane samples.  The orbital period gap is shown between
the cyan lines.}
\end{figure*}

In Fig.\,\ref{excess_period}, we plot the H$\alpha$ EW -- 
estimated as described in Section~\,\ref{ews} -- versus the orbital
period for each CV. As explained in the introduction, short-period CVs
are expected to show stronger H$\alpha$ emission, and this is indeed
confirmed by the figure.

The colour codes and plot symbols are defined in the figure legend and
are used to distinguish between the CV types (symbols) and the emitters
(colours). A small dot in the centre of each symbol refers to the 
data in the CV flag and CV Type column of table~\,\ref{tab1_a}. The
dot is coloured magenta if the CV flag is not equal to zero (the CV is
not in the catalogue or it is of uncertain nature) or the CV type is
of an uncertain nature. If the CV flag is zero and the CV type is
certain, the dot is black. 
   
The short-period quiescent DNe show the greatest EW, and the
survey has no problems in detecting them as emission line
objects. In fact the long-period nova V1425 Aql shows the biggest EW but it misses our initial selection cuts and the EW estimate is too large to be physical, hence it is not shown in the figure.
There are more short-period DNe than long-period DNe in our
sample, which is a characteristic of the Ritter \& Kolb catalogue as a
whole (and expected on evolutionary grounds). 

As already noted in section~\,\ref{orb_per_s}, long-period systems have a
lower H$\alpha$ excess, but the survey is still sensitive enough to
detect many of them as emitters.  As shown in Fig.\,\ref{excess_period} the boundary between systems detected as emitters and those undetected corresponds to a H$\alpha$
EW of $\sim$ 10\AA; the average excess for all systems is clearly
above this value, irrespective of period. Thus most long-period CVs do
exhibit photometrically detectable H$\alpha$ emission, albeit at a 
lower EW level than short-period systems.

There is one object in Fig.\,\ref{excess_period} that shows a large
apparent H$\alpha$ EW, but was not selected as an emitter. This
object is the long-period dwarf nova CW Mon, which is located in a
field of poor data quality. In fact, this field does not meet the 
IPHAS criteria for inclusion in the final survey and will be
re-observed. As a result, the stellar locus exhibits a large scatter
which places the cut further up in the colour-colour plots compared to
fields with good data quality.  The CV is separated from
the locus to the extent that, on our colour-colour plots, it appears above the graphical representations of the selection cuts.  However, it is not
selected due to the large uncertainties of the magnitudes. Past
spectroscopic observations have revealed strong Balmer emission in CW
Mon \citep{1987ApJS...63..685S}.

\subsection{$r^\prime-i^\prime$ vs Orbital Period}

\begin{figure*}
\includegraphics[angle=0,width=\textwidth]{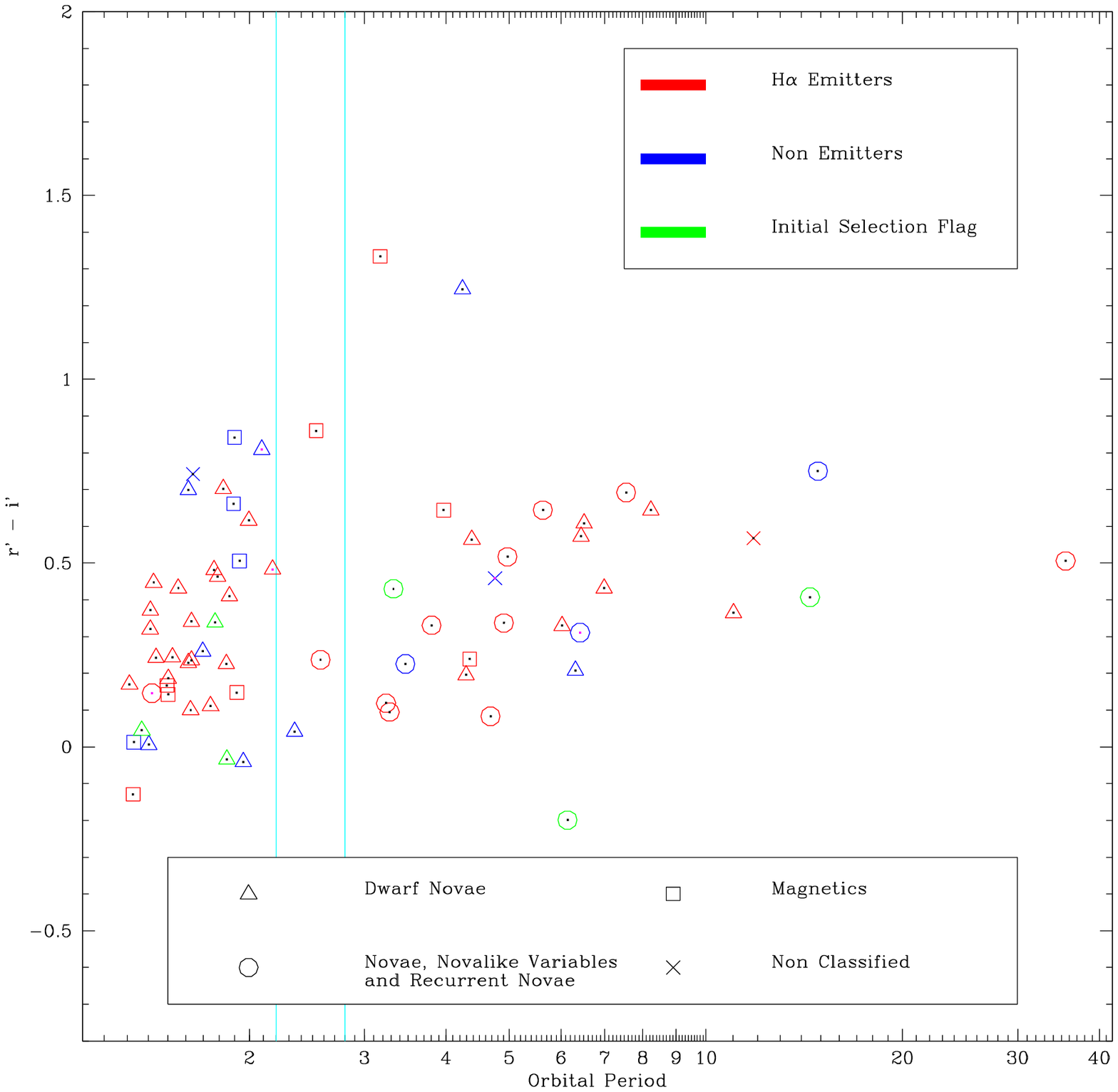}
\caption{\label{rmini_period} A plot of $r^\prime-i^\prime$ versus the
orbital period of CVs in the Off-Plane and In-Plane samples.  The orbital period gap is shown between
the cyan lines.}
\end{figure*}

In Fig.\,\ref{rmini_period} we plot $r^\prime-i^\prime$ colour against orbital
period. No obvious correlation is seen, but we note that the $r^\prime-i^\prime$
colours of the CVs in this plot span a range of more than 1
magnitude. Based on this, we believe the use of selection criteria
based on this broad-band colour (such as a ``blue cut'') would not be
prudent in the construction of a new CV sample from IPHAS. In this
context, it is also worth keeping in mind that the sample of known CVs
analyzed here may already suffer from selection effects. It is
therefore important to minimize the number of {\em a priori} selection
criteria when constructing new samples that are meant to better
represent the true CV parent population.

\subsection{H$\alpha$ EW vs Absolute Magnitude}

\begin{figure*}
\includegraphics[angle=0,width=\textwidth]{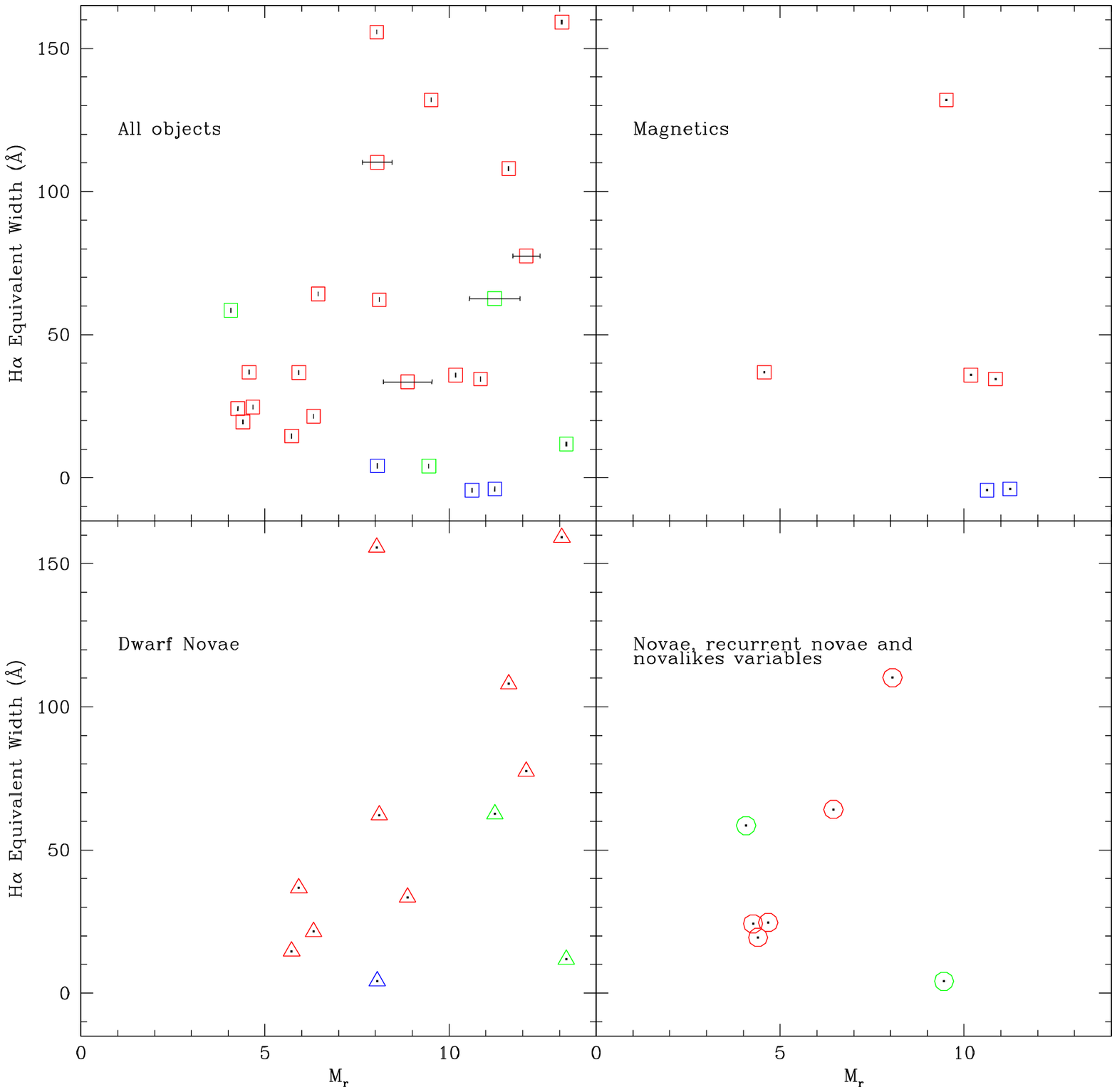}
\caption{\label{mag_ew} The relationship between the absolute
magnitude and the H$\alpha$ equivalent width for the CVs with known
distances.  The symbols are plotted in red if the systems is a H$\alpha$ emitter;
blue if it is a non-emitter; and green if it is a system which did not pass our initial selection cuts.} 
\end{figure*}

In order to test for a correlation between intrinsic brightness and
EW, we calculated absolute magnitudes for CVs with known
distances in our sample by assuming a galactic extinction law of $A_V
= $ 1.5 mag/kpc. In doing so, we also assumed an $R_V = 3.1$
extinction curve to convert $A_V$ to $A_{r^\prime}$.  Fig.\,\ref{mag_ew}
shows the relationship between the resulting absolute magnitudes and
our estimated EWs for various groupings. 

In the plot showing all systems, there appears to be a correlation,
albeit with large scatter. The existence of a correlation becomes
more apparent when we plot the individual CV types separately. For the
DNe and the novae, in particular, there is a definite trend for the
H$\alpha$ EW to increase towards larger (that is fainter) absolute
magnitudes.  The magnetic systems do not show this trend, which is
reasonable given the lack of accretion discs in the AM Her systems. 

Our results agree qualitatively with the work of
\citet{1984ApJS...54..443P}, who has calculated a fit to the
relationship between absolute magnitude and H$\beta$ equivalent 
widths. Note, however, that Patterson removed the contribution of the
secondary from his absolute magnitude estimates and thus considered
only the intrinsic brightness of the accretion disc in his
correlation. We have not made this correction, nor have we corrected
in any way for inclination. Nevertheless, all of our results clearly confirm
that short-period, faint CVs (which are expected to represent
the most numerous CV population) exhibit the strongest H$\alpha$\
emission lines. Thus photometric emission line surveys are indeed an
excellent resource for finding and studying this population.

\section{Discussion}
\label{discuss}

\subsection{Prospects for CV Searches Based on Emission Line Surveys}

A key question we wanted to address with this study was ``can surveys
like IPHAS find members of the postulated large population of faint,
short-period CVs''?  The results here confirm that short-period CVs
generally exhibit relatively large H$\alpha$ excesses in IPHAS, so if 
the hypothesized large population of short-period CVs conform to this
rule, the survey would have no problem detecting them. More
surprisingly, we have also been able to recover long-period CVs as
H$\alpha$ emitters with a comparable success rate. Thus we actually do
not expect any additional bias from the survey in favour of detecting either
long- or short-period CVs. 

The vast majority of short-period systems are believed to be
DNe. However, due to the low mass accretion rates in these systems,
the outburst duty cycle is extremely low. For example, the proto-type
WZ Sge erupts on average every 30~yrs and stays bright for only a few
months. In the context of an emission line survey for CVs, this may
actually be an advantage. The results here show that IPHAS is unable to
detect most DNe in outburst via our selection method, whereas
quiescent DNe tend to show the greatest H$\alpha$ excesses.  Thus
short-period DNe are actually more likely to be discovered by IPHAS if
they have long outburst intervals.

It is interesting to ask how many faint, short-period CVs IPHAS might
eventually discover. Given the low accretion rates and very low mass
secondaries, it is possible that the optical luminosity of these
systems will actually be dominated by their WD primaries. An
accretion-heated WD in a short-period systems would be expected to
have an effective temperature of $T_{eff} \simeq 10000$~K
\citep{2003ApJ...596L.227T}. IPHAS could see such an object out to
275~pc, given its limiting magnitude of $r^\prime=19.5$.  For an assumed
CV space density of $5\times10^{-5}\mathrm{pc^{-3}}$ (see discussion
below), there are roughly $\sim 0.1$ short-period CVs per square degree brighter 
than $r^\prime=19.5$ mags near the galactic plane.  Since IPHAS will
image 1800 square degrees, we would predict that $\sim$ 180 of these
systems could be detected with IPHAS. 

This prediction is conservative in the sense that no active CV should
be fainter than the estimate used above -- any admixture of brighter
CVs will increase the total number considerably (there are
intrinsically fewer of these, but they can be seen to greater
distances). The prediction does, of course, scale directly with the 
assumed space density. The number
adopted above is based on theoretical population synthesis predictions
\citep{1993A&A...271..149K}, and is higher than most existing empirical
estimates \citep{1998PASP..110.1132P, 2005ASPC..330....3G}. This is not a problem, as it is
precisely this discrepancy between theory and observation that an IPHAS CV
sample would be designed to test.

It has already been noted in Section~\,\ref{faint_dependence} that  a
faint CV will on average need to display a greater H$\alpha$ excess
than a bright CV in order to be 
detected. However, if the trend towards larger EWs for short-period, faint DNe
continues to hold for this population, then there may be no problem in
detecting these systems even close to the magnitude limit of the
survey. 

Considering all of the above, it seems clear that IPHAS (and similar
surveys) does have the potential to test theoretical population
synthesis predictions.

\subsection{Implementing an IPHAS-based CV Search}

Initial follow-up to the IPHAS survey has already lead to many clear
H$\alpha$ emitters being spectroscopically observed and
identified. A key problem in constructing a new sample of CVs in this
way is the presence of other large populations of objects 
displaying H$\alpha$ emission. Selecting objects for spectroscopic follow-up
using the selection techniques illustrated here can uncover most of
the CVs, but the fraction of CVs amongst objects 
selected in this way may not be large. Our results to
date suggest that early type emission line stars (for example Be stars)
tend to dominate bright ($r^\prime \ltappeq 18$) samples of H$\alpha$ excess
objects. Thus many non-CVs need to be observed in the process of
constructing a new sample of CVs, unless the distribution of H$\alpha$
excess objects changes significantly towards fainter magnitudes. We
are currently investigating this possibility.

\section{Conclusions}
 \label{conc}

We have considered the properties of known CVs contained within
the galactic plane H$\alpha$ survey IPHAS as of July 2004, along with those of a
sample of CVs outside the galactic plane observed with the same
observational setup.  We have developed routines to 
select clear H$\alpha$ emitters from the IPHAS photometry and used 
these routines to determine the percentage of CVs in our samples that
would be recovered as H$\alpha$ emitters.  

We find that the overall completeness of IPHAS-like surveys for
detecting CVs as emission line objects is $\simeq$~70 per cent,
roughly independent of CV type and orbital period. Our recovery
fractions are steady near this value up to at least $r^\prime \simeq
18.5$ and only drop towards the very faintest magnitudes we consider
($r^\prime = 19.5$). There are 23 CVs in our samples not recovered as H$\alpha$ emitters. Six do not pass our initial selection cuts, and 2 are without a known sub-type. Several are not recovered due to observing some CVs when H$\alpha$ emission is expected to be weak or absent, for example DNe in outburst, and magnetic CVs in a low state.  The remainder of non-recovered CVs are due to scatter in faint magnitude bins and cases where the fitting process did not produce good results.

We have estimated H$\alpha$ EWs from our photometry 
and find that, as expected, short-period CVs show the strongest
H$\alpha$ emission lines. This is encouraging for the prospects of
finding the predicted population of faint, short-period
CVs. However, we also find that most long-period CVs exhibit strong
enough H$\alpha$\ emission for them to be easily detected by
IPHAS-like surveys.  

In closing, we expect that IPHAS (and related surveys) will allow us
to construct samples of CVs that are free from most of the
selection effects that have plagued previous comparisons between
theoretical population synthesis predictions and observations. In
particular, such samples should uncover the long-predicted large
population of faint, short-period systems, if it exists.

\section*{Acknowledgments}
Based in part
on observations made the Isaac Newton Telescope, which is operated on
the island of La Palma by the Isaac Newton Group in the Spanish
Observatorio del Roque de los Muchachos of the IAC.  ARW was supported by a
PPARC Studentship. BTG was supported by a PPARC Advanced Fellowship.   DS acknowledges a
Smithsonian Astrophysical Observatory Clay Fellowship.
\bibliographystyle{mn_new}
\bibliography{mn-jour,arwbib}

\bsp

\label{lastpage}

\end{document}